%% file: main.tex
\pdfoutput=1
\documentclass[12pt,a4paper]{article}

\usepackage{ifthen} 
\newboolean{pdflatex}
\setboolean{pdflatex}{true} 

\newboolean{articletitles}
\setboolean{articletitles}{true} 

\newboolean{uprightparticles}
\setboolean{uprightparticles}{false} 

\def\paperauthors{LHCb collaboration} 
\def\paperasciititle{Measurement of the W-boson production cross-sections in $pp$ collisions at sqrt(s) = 13 TeV with the LHCb detector} 
\def\papertitle{Measurement of the \PW-boson production cross-sections in $pp$ collisions at
$\sqs = 13\tev$ in the forward region} 
\def\paperkeywords{{High Energy Physics}, {LHCb}} 
\def\papercopyright{\the\year\ CERN for the benefit of the LHCb collaboration} 
\def\paperlicence{CC BY 4.0 licence}
\def\paperlicenceurl{https://creativecommons.org/licenses/by/4.0/}

\def\eTrk     {{\ensuremath{\varepsilon_{\textrm{trk}}}}\xspace}
\def\eID      {{\ensuremath{\varepsilon_{\textrm{ID $\vert$ trk}}}}\xspace}
\def\eLZero   {{\ensuremath{\varepsilon_{\textrm{trig $\vert$ ID}}}}\xspace}

\def\ZToMuMu     {\decay{Z}{\mup \mun}}
\def\WToTauNu     {\decay{W}{\tau \nu}}
\def\WToMuNu     {\decay{W}{\mu \nu}}
\def\WpToMuNu     {\decay{\Wp}{\mup \neum}}
\def\WmToMuNu     {\decay{\Wm}{\mun \neumb}}

\def\UpsilonToMuMu     {\decay{\OneS}{\mup \mun}}

\newif\ifEnableSectionTOCLinks
\EnableSectionTOCLinksfalse 

\input{LHCb/preamble}
\usepackage{longtable} 
\usepackage{rotating}

\begin{document}

\renewcommand{\thefootnote}{\fnsymbol{footnote}}
\setcounter{footnote}{1}

\input{title-LHCb-PAPER}


\renewcommand{\thefootnote}{\arabic{footnote}}
\setcounter{footnote}{0}


\cleardoublepage


\pagestyle{plain} 
\setcounter{page}{1}
\pagenumbering{arabic}


\input{body}

\input{LHCb/acknowledgements}

\clearpage


\clearpage


\addcontentsline{toc}{section}{References}
\bibliographystyle{LHCb/LHCb}
\bibliography{main,LHCb/standard,LHCb/LHCb-PAPER,LHCb/LHCb-CONF,LHCb/LHCb-DP,LHCb/LHCb-TDR,LHCb/LHCb-PUB}

\newpage
\input{Authorship_LHCb-PAPER-2025-070.tex}

%
%
%
%

\end{document}

%% file: LHCb/preamble.tex
\usepackage[top=1in, bottom=1.25in, left=1in, right=1in]{geometry}

\usepackage{microtype}
\usepackage{lineno}  
\usepackage{xspace} 
\usepackage{caption} 

\usepackage{graphicx}  
\usepackage{color}
\usepackage{colortbl}
\graphicspath{{./figs/}} 

\usepackage{amsmath} 
\usepackage{amssymb}
\usepackage{amsfonts}
\usepackage{upgreek} 

\newcommand*\patchAmsMathEnvironmentForLineno[1]{%
\expandafter\let\csname old#1\expandafter\endcsname\csname #1\endcsname
\expandafter\let\csname oldend#1\expandafter\endcsname\csname
end#1\endcsname
 \renewenvironment{#1}%
   {\linenomath\csname old#1\endcsname}%
   {\csname oldend#1\endcsname\endlinenomath}%
}
\newcommand*\patchBothAmsMathEnvironmentsForLineno[1]{%
  \patchAmsMathEnvironmentForLineno{#1}%
  \patchAmsMathEnvironmentForLineno{#1*}%
}
\AtBeginDocument{%
\patchBothAmsMathEnvironmentsForLineno{equation}%
\patchBothAmsMathEnvironmentsForLineno{align}%
\patchBothAmsMathEnvironmentsForLineno{flalign}%
\patchBothAmsMathEnvironmentsForLineno{alignat}%
\patchBothAmsMathEnvironmentsForLineno{gather}%
\patchBothAmsMathEnvironmentsForLineno{multline}%
\patchBothAmsMathEnvironmentsForLineno{eqnarray}%
}

\usepackage[pdftex,
            pdfauthor={\paperauthors},
            pdftitle={\paperasciititle},
            pdfkeywords={\paperkeywords}]{hyperref}
\usepackage{hyperxmp}
\hypersetup{
    pdfcopyright={Copyright (C) \papercopyright},
    pdflicenseurl={\paperlicenceurl}
}

\usepackage[colorinlistoftodos,textsize=scriptsize]{todonotes}

\usepackage[bottom,flushmargin,hang,multiple]{footmisc}

\usepackage[all]{hypcap} 

\input{LHCb/lhcb-symbols-def} 

\hypersetup{
  colorlinks   = true, 
  urlcolor     = blue, 
  linkcolor    = blue, 
  citecolor    = red   
}

\ifEnableSectionTOCLinks
    \usepackage[explicit]{titlesec} 
    
    \let\oldcontentsline\contentsline
    \renewcommand

    \titleformat{\section}{\normalfont\Large\bf}{\hyperlink{tocsection.\thesection}{{\thesection} \parbox[t]{\dimexpr\textwidth-1pc}{#1}}}{1pc}{}

    \titleformat{\subsection}{\normalfont\bf}{\hyperlink{tocsubsection.\thesubsection}{{\thesubsection} \parbox[t]{\dimexpr\textwidth-1pc}{#1}}}{1pc}{}

    \titleformat{name=\section,numberless}[display]{}{}{0pt}{\normalfont\Huge\bfseries #1}
\fi

\usepackage{cite} 
\usepackage{LHCb/mciteplus}

%% file: LHCb/lhcb-symbols-def.tex
\usepackage{xspace} 
\usepackage{upgreek}

\def\lhcb   {\mbox{LHCb}\xspace}
\def\atlas  {\mbox{ATLAS}\xspace}
\def\cms    {\mbox{CMS}\xspace}
\def\alice  {\mbox{ALICE}\xspace}

\def\MagUp {\mbox{\em Mag\kern -0.05em Up}\xspace}

\ifthenelse{\boolean{uprightparticles}}%
{

 \def\Pmu         {\ensuremath{\upmu}\xspace}                 
 \def\Pnu         {\ensuremath{\upnu}\xspace}

 \def\Ppsi        {\ensuremath{\uppsi}\xspace}

 \def\PDelta      {\ensuremath{\Delta}\xspace}                 
 \def\PXi         {\ensuremath{\Xi}\xspace}                 
 \def\PLambda     {\ensuremath{\Lambda}\xspace}                 
 \def\PSigma      {\ensuremath{\Sigma}\xspace}                 
 \def\POmega      {\ensuremath{\Omega}\xspace}                 
 \def\PUpsilon    {\ensuremath{\Upsilon}\xspace}
 \let\oldPi\Pi
 \def\PPi         {\ensuremath{\oldPi}\xspace}

 \def\PB      {\ensuremath{\mathrm{B}}\xspace}                 
 \def\PD      {\ensuremath{\mathrm{D}}\xspace}                 
 \def\PJ      {\ensuremath{\mathrm{J}}\xspace}                 
 \def\PK      {\ensuremath{\mathrm{K}}\xspace}                 
 \def\PW      {\ensuremath{\mathrm{W}}\xspace}                 
 \def\PZ      {\ensuremath{\mathrm{Z}}\xspace}                 
 \def\Pb      {\ensuremath{\mathrm{b}}\xspace}                 
 \def\Pc      {\ensuremath{\mathrm{c}}\xspace}

 \def\Ps      {\ensuremath{\mathrm{s}}\xspace}                 
 \def\Pt      {\ensuremath{\mathrm{t}}\xspace}                 

 \def\thebaroffset{0.0em}
}
{

 \def\Pmu         {\ensuremath{\mu}\xspace}                 
 \def\Pnu         {\ensuremath{\nu}\xspace}

 \def\Ppsi        {\ensuremath{\psi}\xspace}                 
                  
 \mathchardef\PDelta="7101
 \mathchardef\PXi="7104
 \mathchardef\PLambda="7103
 \mathchardef\PSigma="7106
 \mathchardef\POmega="710A
 \mathchardef\PUpsilon="7107
 \mathchardef\PPi="7105
 \def\PB      {\ensuremath{B}\xspace}                 
 \def\PD      {\ensuremath{D}\xspace}                 
 \def\PJ      {\ensuremath{J}\xspace}                 
 \def\PK      {\ensuremath{K}\xspace}                 
 \def\PW      {\ensuremath{W}\xspace}                 
 \def\PZ      {\ensuremath{Z}\xspace}                 
 \def\Pb      {\ensuremath{b}\xspace}                 
 \def\Pc      {\ensuremath{c}\xspace}

 \def\Ps      {\ensuremath{s}\xspace}                 
 \def\Pt      {\ensuremath{t}\xspace}                 

 \def\thebaroffset{0.18em}
}
\newcommand{\offsetoverline}[2][\thebaroffset]{\kern #1\overline{\kern -#1 #2}}%

\makeatletter
\ifcase \@ptsize \relax
  \newcommand{\miniscule}{\@setfontsize\miniscule{4}{5}}
\or
  \newcommand{\miniscule}{\@setfontsize\miniscule{5}{6}}
\or
  \newcommand{\miniscule}{\@setfontsize\miniscule{5}{6}}
\fi
\makeatother

\DeclareRobustCommand{\optbar}[1]{\shortstack{{\miniscule (\rule[.5ex]{1.25em}{.18mm})}
  \\ [-.7ex] $#1$}}

\def\mup        {{\ensuremath{\Pmu^+}}\xspace}
\def\mun        {{\ensuremath{\Pmu^-}}\xspace} 
\def\mupm       {{\ensuremath{\Pmu^\pm}}\xspace}

\def\neu        {{\ensuremath{\Pnu}}\xspace}
\def\neub       {{\ensuremath{\overline{\Pnu}}}\xspace}
\def\neum       {{\ensuremath{\neu_\mu}}\xspace}
\def\neumb      {{\ensuremath{\neub_\mu}}\xspace}

\def\W      {{\ensuremath{\PW}}\xspace}
\def\Wp     {{\ensuremath{\PW^+}}\xspace}
\def\Wm     {{\ensuremath{\PW^-}}\xspace}
\def\Wpm    {{\ensuremath{\PW^\pm}}\xspace}
\def\Z      {{\ensuremath{\PZ}}\xspace}

\def\squark    {{\ensuremath{\Ps}}\xspace}

\def\cquark    {{\ensuremath{\Pc}}\xspace}
\def\cquarkbar {{\ensuremath{\overline \cquark}}\xspace}
\def\ccbar     {{\ensuremath{\cquark\cquarkbar}}\xspace}
\def\bquark    {{\ensuremath{\Pb}}\xspace}
\def\bquarkbar {{\ensuremath{\overline \bquark}}\xspace}
\def\bbbar     {{\ensuremath{\bquark\bquarkbar}}\xspace}
\def\tquark    {{\ensuremath{\Pt}}\xspace}
\def\tquarkbar {{\ensuremath{\overline \tquark}}\xspace}
\def\ttbar     {{\ensuremath{\tquark\tquarkbar}}\xspace}

\def\KorKbar {\kern \thebaroffset\optbar{\kern -\thebaroffset \PK}{}\xspace}

\def\D       {{\ensuremath{\PD}}\xspace}

\def\DorDbar {\kern \thebaroffset\optbar{\kern -\thebaroffset \PD}\xspace}

\def\Dp      {{\ensuremath{\D^+}}\xspace}
\def\Dm      {{\ensuremath{\D^-}}\xspace}

\def\DpDm    {\ensuremath{\Dp {\kern -0.16em \Dm}}\xspace}

\def\B       {{\ensuremath{\PB}}\xspace}

\def\BorBbar {\kern \thebaroffset\optbar{\kern -\thebaroffset \PB}\xspace}

\def\Bd      {{\ensuremath{\B^0}}\xspace}

\def\BdorBdbar {\kern \thebaroffset\optbar{\kern -\thebaroffset \Bd}\xspace}

\def\Bs      {{\ensuremath{\B^0_\squark}}\xspace}

\def\BsorBsbar {\kern \thebaroffset\optbar{\kern -\thebaroffset \Bs}\xspace}

\def\jpsi     {{\ensuremath{{\PJ\mskip -3mu/\mskip -2mu\Ppsi}}}\xspace}

\def\Y#1S{\ensuremath{\PUpsilon{(#1S)}}\xspace}
\def\OneS  {{\Y1S}\xspace}

\def\LorLbar     {\kern \thebaroffset\optbar{\kern -\thebaroffset \PLambda}\xspace}

\newcommand{\decay}[2]{\ensuremath{\mathinner{#1\!\to #2}}\xspace}

\def\to                 {\ensuremath{\rightarrow}\xspace}

\def\AT#1     {\ensuremath{A_{\mathrm{T}}^{#1}}\xspace}           

\def\C#1      {\ensuremath{\mathcal{C}_{#1}}\xspace}                       
\def\Cp#1     {\ensuremath{\mathcal{C}_{#1}^{'}}\xspace}                    
\def\Ceff#1   {\ensuremath{\mathcal{C}_{#1}^{\mathrm{(eff)}}}\xspace}        
\def\Cpeff#1  {\ensuremath{\mathcal{C}_{#1}^{'\mathrm{(eff)}}}\xspace}       
\def\Ope#1    {\ensuremath{\mathcal{O}_{#1}}\xspace}                       
\def\Opep#1   {\ensuremath{\mathcal{O}_{#1}^{'}}\xspace}                    

\newcommand{\nospaceunit}[1]{\ensuremath{\text{#1}}}       
\newcommand{\aunit}[1]{\ensuremath{\text{\,#1}}}       

\newcommand{\tev}{\aunit{Te\kern -0.1em V}\xspace}
\newcommand{\gev}{\aunit{Ge\kern -0.1em V}\xspace}
\newcommand{\mev}{\aunit{Me\kern -0.1em V}\xspace}
\newcommand{\kev}{\aunit{ke\kern -0.1em V}\xspace}
\newcommand{\ev}{\aunit{e\kern -0.1em V}\xspace}
 
\newcommand{\mevc}{\ensuremath{\aunit{Me\kern -0.1em V\!/}c}\xspace}
\newcommand{\gevc}{\ensuremath{\aunit{Ge\kern -0.1em V\!/}c}\xspace}
\newcommand{\mevcc}{\ensuremath{\aunit{Me\kern -0.1em V\!/}c^2}\xspace}
\newcommand{\gevcc}{\ensuremath{\aunit{Ge\kern -0.1em V\!/}c^2}\xspace}

\def\mum  {\ensuremath{\,\upmu\nospaceunit{m}}\xspace}

\def\fb   {\ensuremath{\aunit{fb}}\xspace}
\def\invfb   {\ensuremath{\fb^{-1}}\xspace}

\def\deriv {\ensuremath{\mathrm{d}}}

\def\gsim{{~\raise.15em\hbox{$>$}\kern-.85em
          \lower.35em\hbox{$\sim$}~}\xspace}
\def\lsim{{~\raise.15em\hbox{$<$}\kern-.85em
          \lower.35em\hbox{$\sim$}~}\xspace}

\def\sqs   {\ensuremath{\protect\sqrt{s}}\xspace}

\def\pt         {\ensuremath{p_{\mathrm{T}}}\xspace}

\def\ptot       {\ensuremath{p}\xspace}

\def\gauss      {\mbox{\textsc{Gauss}}\xspace}
\def\geant      {\mbox{\textsc{Geant4}}\xspace}

\def\photos     {\mbox{\textsc{Photos}}\xspace}
\def\powheg     {\mbox{\textsc{Powheg}}\xspace}
\def\pythia     {\mbox{\textsc{Pythia}}\xspace}
\def\resbos     {\mbox{\textsc{ResBos}}\xspace}

\def\tell1  {TELL1\xspace}
\def\ukl1   {UKL1\xspace}

\newcommand{\vs}{\mbox{\itshape vs.}\xspace}
\newcommand{\phz}{\phantom{0}}
\newcommand{\lhcborcid}[1]{\href{https://orcid.org/#1}{\hspace*{0.1em}\raisebox{-0.45ex}{\includegraphics[width=1em]{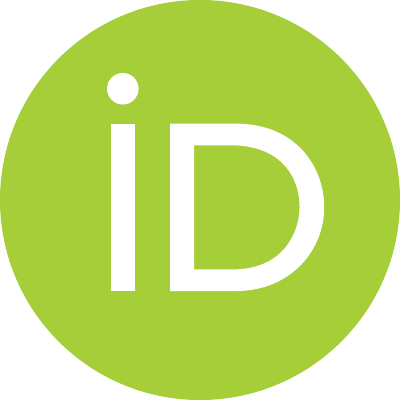}}}}


%% file: title-LHCb-PAPER.tex

\begin{titlepage}
\pagenumbering{roman}

\vspace*{-1.5cm}
\centerline{\large EUROPEAN ORGANIZATION FOR NUCLEAR RESEARCH (CERN)}
\vspace*{1.5cm}
\noindent
\begin{tabular*}{\linewidth}{lc@{\extracolsep{\fill}}r@{\extracolsep{0pt}}}
\ifthenelse{\boolean{pdflatex}}
{\vspace*{-1.5cm}\mbox{\!\!\!\includegraphics[width=.14\textwidth]{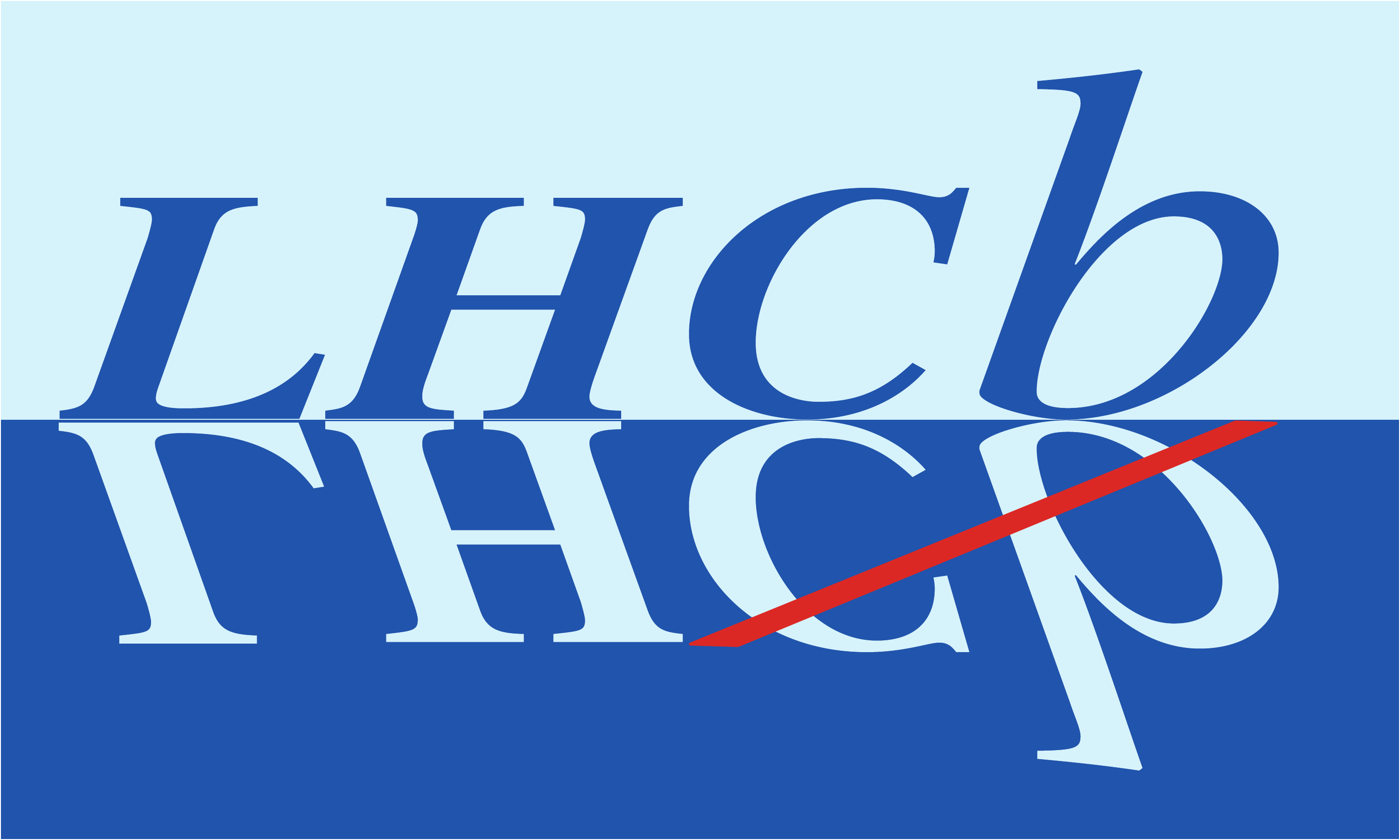}} & &}%
{\vspace*{-1.2cm}\mbox{\!\!\!\includegraphics[width=.12\textwidth]{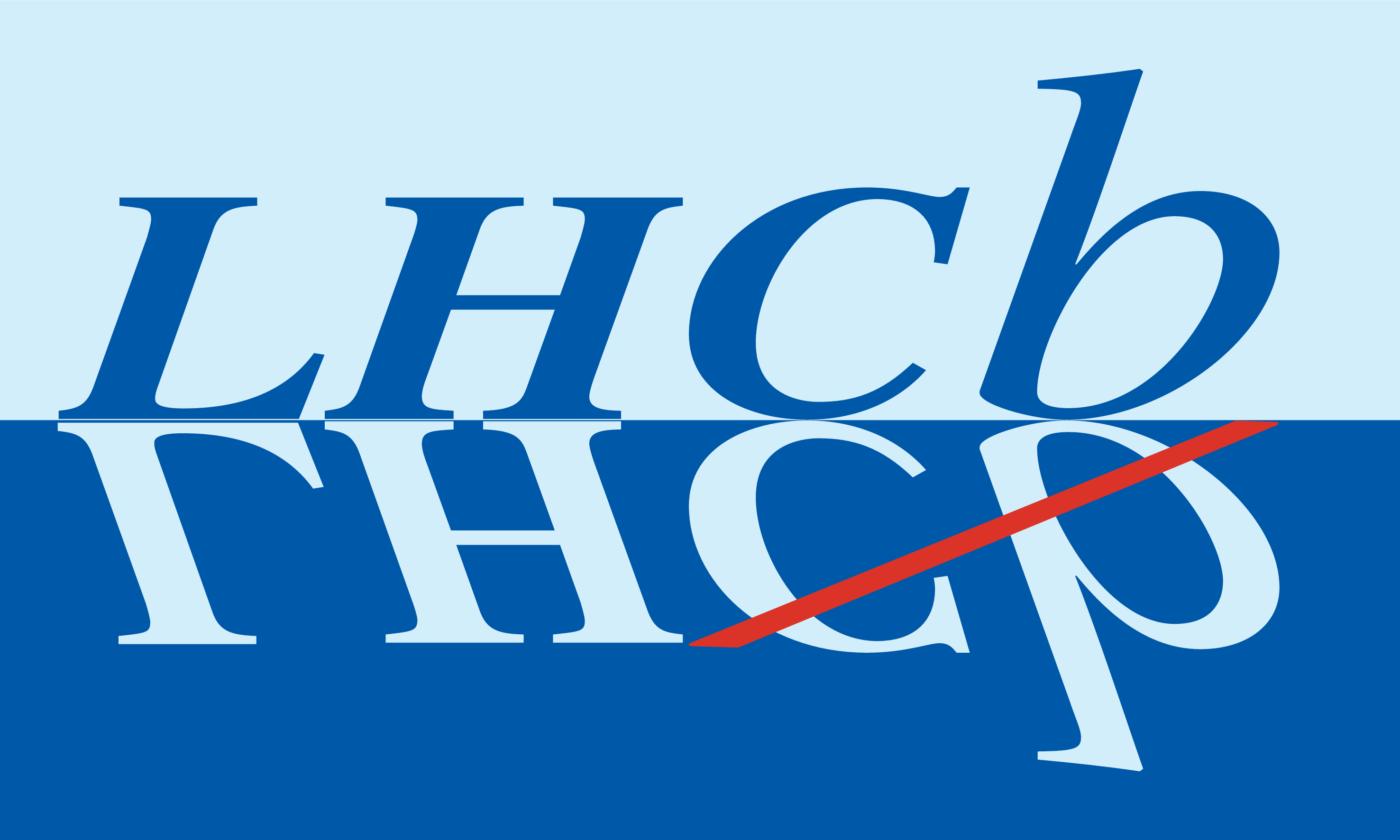}} & &}%
\\
 & & CERN-EP-2026-083 \\  
 & & LHCb-PAPER-2025-070 \\  
 & & April 14, 2026 \\ 
 & & \\
\end{tabular*}

\vspace*{2.5cm}

{\normalfont\bfseries\boldmath\huge
\begin{center}
  \papertitle 
\end{center}
}

\vspace*{1.5cm}

\begin{center}
\paperauthors\footnote{Authors are listed at the end of this paper.}
\end{center}

\vspace{\fill}

\begin{abstract}
  \noindent
A precision measurement of the $W$-boson production cross-section is performed using the $W \to \mu\nu$ decay channel, based on a sample of proton-proton collision data collected by the \lhcb experiment at $\sqs = 13\tev$ and corresponding to an integrated luminosity of 5.1\invfb.
The cross-section is measured for muons with transverse momentum between 25 and 55\gev and pseudorapidity between 2.0 and 4.5. 
The integrated production cross-sections of $W$ bosons are measured to be
\begin{equation*}
\begin{aligned}
  \sigma_{W^+ \to \mu^+\nu} &= 1754.2 \pm 1.5 \pm 11.9 \pm 35.1\text{ pb}\\
  \sigma_{W^- \to \mu^-\overline{\nu}} &= 1178.1 \pm 1.3 \pm \phz9.7 \pm 23.6\text{ pb}\\
\end{aligned}
\end{equation*}
where uncertainties are statistical, systematic, and due to the luminosity determination, respectively. Results are in good agreement with theoretical predictions at next-to-next-to-leading order in perturbative quantum chromodynamics. This measurement is significantly more precise than previous results in this kinematic regime.
\end{abstract}

\vspace*{2.0cm}

\begin{center}
  Published in
  Phys.~Rev.~D 114, 052003 (2026)
\end{center}

\vspace{\fill}

{\footnotesize 
\centerline{\copyright~\papercopyright. \href{\paperlicenceurl}{\paperlicence}.}}
\vspace*{2mm}

\end{titlepage}


\newpage
\setcounter{page}{2}
\mbox{~}
%
%
%
%

%% file: body.tex
\section{Introduction} 
Proton parton distribution functions (PDFs) are a fundamental input for cross-section calculations at hadron colliders, and frequently represent the dominant source of systematic uncertainty in precision measurements~\cite{LHCb-PAPER-2021-024,LHCb-PAPER-2024-028}. As nonperturbative quantities are not directly calculable within the Standard Model (SM), PDFs must be extracted empirically through global analyses using experimental inputs, including a wide range of scattering processes~\cite{H1:2015ubc,H1:2004esl,Moreno:1990sf,NuSea:2001idv,H1:2010fzx}. Measurements of the electroweak (\W and \Z) boson production cross-sections at Large Hadron Collider energies provide not only a stringent test of the SM, but also a critical constraint for global fits of proton PDFs~\cite{Deng:2020sol}. 

Theoretical predictions for electroweak production cross-sections, calculated up to next-to-next-to-next-to-leading order (N$^3$LO) in perturbative quantum chromodynamics (QCD)~\cite{PhysRevD.51.44, Hamberg:1990np,vanNeerven:1991gh,PhysRevLett.88.201801, Anastasiou:2003ds,Ju:2021lah,Alioli:2021qbf,Ebert:2020dfc,Bizon:2019zgf,Billis:2024dqq}, critically depend on the parameterization of Bjorken-$x$ distributions within the proton. Here, $x$ represents the longitudinal momentum fraction carried by constituent partons~\cite{Brodsky:2016fyh}, serving as a fundamental parameter that governs the nonperturbative QCD dynamics across different energy scales.
The \lhcb experiment's forward acceptance ($2 < \eta < 5$) uniquely probes kinematic regimes that complement measurements from \atlas~\cite{ATLAS:2011qdp,ATLAS:2024nrd,ATLAS:2014alx,ATLAS:2015iiu,ATLAS:2019zci,ATLAS:2019fyu,ATLAS:2018pyl,ATLAS:2016nqi,ATLAS:2016fij,ATLAS:2024irg}, \cms~\cite{PhysRevD.102.092012,CMS:2011aa,CMS:2014pkt,CMS:2015hyl,CMS:2011wyd,CMS:2022ubq,CMS:2014dyj,CMS:2024myi,CMS:2010svw} and \alice~\cite{Sakai:2025xnd, ALICE:2022cxs, Valle:2021skk, Stocco:2017inb, ALICE:2016rzo, Buthelezi:2015cbr, Senosi:2015omk}, particularly in both the large and small-$x$ regions that remain theoretically challenging. 
For typical $W$-boson production kinematics in the \lhcb acceptance, the corresponding parton momentum fractions span approximately $10^{-4}$ to $10^{-1}$, providing complementary coverage to that of other experiments.

The \lhcb collaboration has previously published comprehensive measurements of inclusive $W$- and $Z$-boson production in 
proton-proton ($pp$) collisions at center-of-mass energies of 
$\sqs=7$, 8, and 13\tev~\cite{LHCb-PAPER-2012-008,LHCb-PAPER-2012-036,LHCb-PAPER-2014-033,LHCb-PAPER-2015-003,LHCb-PAPER-2015-001,LHCb-PAPER-2015-049,LHCb-PAPER-2016-021,LHCb-PAPER-2016-024,LHCb-PAPER-2021-037,LHCb-PAPER-2023-010} with integrated luminosities of approximately 1, 2, and 1.6\invfb respectively.
The present analysis utilizes the Run~2 dataset collected during 2016--2018, corresponding to an integrated luminosity of 5.1\invfb. Precise measurements of inclusive \Wp and \Wm-boson production cross-sections are presented for muons in the fiducial region with transverse momentum $\pt^{\mu}$ between $25$ and $55\gev$\footnote{Natural units with $\hbar=c=1$ are used throughout.} and pseudorapidity $\eta_{\mu}$ between 2.0 and 4.5, complemented by differential cross-section measurements binned in muon pseudorapidity. In addition, a measurement of the muon charge asymmetry, using the same sample of \W-boson decays, is presented in a companion article~\cite{LHCb:2026biq}.

The differential $W$-boson production cross-section is measured in 18 $\eta_{\mu}$ intervals, according to
 \begin{equation}
   \frac{\deriv \sigma_{W^{\pm} \to \mu^{\pm}\nu}}{\deriv \eta^{\mu^{\pm}}} = \frac{N^{\pm}_{\rm yield} \cdot f^{\Wpm}_{\rm FSR}}{ \Delta\eta^{\mu^{\pm}} \cdot \mathcal{L} \cdot \varepsilon_{W^{\pm}}},
   \label{eq:cross-section}
 \end{equation}   
where $\Delta\eta^{\mu^{\pm}}$ is the size of the interval, $\mathcal{L}$ the integrated luminosity, $N^{\pm}_{\rm yield}$ the \Wpm candidate yield determined from a template fit to the muon \pt spectrum, $\varepsilon_{W^{\pm}}$  the total reconstruction and selection efficiency, and $f^{\Wpm}_{\rm FSR}$ the final-state radiation (FSR) correction. 

Born-level QED corrections are applied to the measured cross-section for direct comparison with theoretical predictions: final-state radiation corrections are applied using a ratio of \resbos2~\cite{PhysRevD.86.094026} predictions with and without \photos~\cite{davidson2015photos} implementation, which maps bare-level muon kinematics to Born-level definitions.

\section{Detector and simulation}
The \lhcb detector~\cite{LHCb-DP-2008-001,LHCb-DP-2014-002} is a single-arm forward
spectrometer covering the \mbox{pseudorapidity} range $2<\eta <5$,
designed for the study of particles containing \bquark or \cquark
quarks. The detector used to collect the data analyzed in this paper includes a high-precision tracking system
consisting of a silicon-strip vertex detector surrounding the $pp$
interaction region~\cite{LHCb-DP-2014-001}, a large-area silicon-strip detector located
upstream of a dipole magnet with a bending power of about
$4{\mathrm{\,T\,m}}$, and three stations of silicon-strip detectors and straw
drift tubes~\cite{LHCb-DP-2017-001} placed downstream of the magnet.
The tracking system provides a measurement of the momentum, \ptot, of charged particles with
a relative uncertainty that varies from 0.5\% at low momentum to 1.0\% at 200\gev.
The minimum distance of a track to a primary $pp$ collision vertex (PV), the impact parameter (IP), 
is measured with a resolution of $(15+29/\pt)\mum$,
where \pt is the component of the momentum transverse to the beam, in\,\gev.
The magnetic field deflects oppositely charged particles in opposite directions and this can lead to detection asymmetries. Periodically reversing the magnetic field polarity throughout the data taking almost cancels the effect.
Different types of charged hadrons are distinguished using information
from two ring-imaging Cherenkov detectors~\cite{LHCb-DP-2012-003}. 
Photons, electrons and hadrons are identified by a calorimeter system consisting of
scintillating-pad and preshower detectors, an electromagnetic
and a hadronic calorimeter. Muons are identified by a
system composed of alternating layers of iron and multiwire
proportional chambers~\cite{LHCb-DP-2012-002}.

The online event selection is performed by a trigger~\cite{LHCb-DP-2012-004}, 
which consists of a hardware stage based on information from the calorimeter and muon
systems, followed by a software stage, which applies a full event
reconstruction.
Triggered data further undergo a centralised, offline processing step
to deliver physics-analysis-ready data across the entire \lhcb physics programme~\cite{Stripping}.

Simulation is required to model the effects of the detector acceptance and the imposed selection requirements, and to describe the signal and background components when fitting data to extract yields.
In the simulation, $pp$ collisions are generated using
\pythia~\cite{Sjostrand:2007gs} with a specific \lhcb configuration~\cite{LHCb-PROC-2010-056}, in which FSR is generated using \photos.
The interaction of the generated particles with the detector, and its response, are implemented using the \geant toolkit~\cite{Allison:2006ve, Agostinelli:2002hh} as described in Ref.~\cite{LHCb-PROC-2011-006}. To evaluate higher-order effects in the perturbative expansion, this analysis employs multiple event generators. \textsc{DYTurbo}~\cite{DYTurbo2020} provides high-precision QCD \pt resummation for Drell--Yan cross-sections, achieving approximate next-to-next-to-next-to-next-to-leading logarithmic (N$^4$LL) accuracy and serving as the weighting tool described in Sec.~\ref{subsec:correction_sim}.
For the background cross-section predictions described in Sec.~\ref{sec:signal_yield}, MCFM~\cite{MCFM}, a parton-level event generator that gives predictions for a wide range of processes at hadron colliders, is used.
\resbos~2 and \powheg~\cite{POWHEGBoxV2SingleBoson} are utilized to obtain theoretical predictions for experimental results.
\resbos2 integrates a next-to-next-to-leading order‌ (NNLO) fixed-order calculation for high-\pt bosons with the Collins--Soper--Sterman resummation formalism~\cite{COLLINS1985199, COLLINS1981381, COLLINS1982446} at low \pt, implementing an all-orders resummation of large logarithmic terms from gluon emissions. The \powheg generator simulates the hard interaction at NLO in perturbative QCD. For the \WToMuNu process, all predictions refer to the fiducial region mentioned above. 

\section{Event selection}
The signal selection employs a multi-stage optimization strategy that balances signal purity against reconstruction efficiency. 
Muon candidates must satisfy stringent quality criteria including good track-fit quality, a precise momentum measurement ($\sigma(p)/p < 0.06$), physical momentum limits ($p < 2\tev$) and satisfy the fiducial criteria. 

Additional selection requirements are imposed to reject specific backgrounds. Nonprompt muons from heavy-flavor decays are rejected by requiring less than 5.0\gev energy deposition in the hadronic calorimeter cells surrounding the muon trajectory, combined with muon identification (ID) criteria.
Track isolation requirements ($\sum p_{\rm T}^{\text{track}} < 3.0\gev$ in a cone around the muon within \mbox{$\sqrt{(\phi-\phi_\mu)^2 + (\eta - \eta_\mu)^2} < 0.5$}, where $\phi$ is the azimuthal angle) effectively suppress hadronic backgrounds.
A \Z-boson veto is also added to reject events containing a second muon with opposite charge, which could form \ZToMuMu candidates ($\pt^{\prime} > 25\gev$, $2.0 < \eta^{\prime} < 4.5$, where $\pt^{\prime}$ and $\eta^{\prime}$ are properties of the second muon).  
Tight impact parameter significance requirements further reduce contributions from \WToTauNu and heavy-flavor decays. 
After applying these selection criteria, around $6.3 \times 10^6$ ($4.4 \times 10^6$) \Wp (\Wm) candidates are retained with an approximate purity of 85\% (80\%).

\section{Corrections}
To account for detector misalignment effects, imperfections in the simulation of the detector response and resolution, and missing higher-order effects in theoretical predictions, dedicated calibration and correction procedures are implemented for both data and simulation.

\subsection{Data calibrations}
The charged particle momentum measurements are affected by variations in detector conditions throughout the course of the data-taking period.
These effects are suppressed using \UpsilonToMuMu decays in data, where the reconstructed \OneS mass is corrected to the known value~\cite{PDG2024} for each year of data-taking.
The mass distribution of the \OneS peak is described by the sum of two Crystal Ball functions~\cite{Skwarnicki:1986xj} with a common peak position, while the background is described by an exponential function. 
The Crystal Ball peak positions $M_{\rm CB}$ are then compared to the known \OneS mass $M_{\rm PDG}$~\cite{PDG2024} to determine the momentum scale correction $S_\alpha = M_{\rm PDG}/M_{\rm CB}$.

For remaining misalignment effects affecting high-\pt muons, the pseudomass method is employed~\cite{LHCb-DP-2023-001,Barter:2021npd}. This technique estimates two-particle final-state masses in the absence of full information on the final-state momenta. For \ZToMuMu decays, pseudomasses are defined as
\begin{equation}
\mathcal{M}^{\pm} \equiv \sqrt{2p^{\pm}p_{\rm{T}}^{\pm}\frac{p^{\mp}}{p_{\rm{T}}^{\mp}}\left(1 - \cos\theta\right)},
\end{equation} 
where $p^{\pm}$ ($\pt^{\pm}$) are the \mupm (transverse) momenta, and $\theta$ is the opening angle between the two muons.
Unlike the invariant mass, where biases cancel between \mup and \mun tracks, the pseudomasses maintain first-order sensitivity to curvature biases
\begin{equation}
\label{eq:bias}                                                    \frac{q}{p} \to  \frac{q}{S_\alpha p} + \delta, \end{equation}
where $q$ is the muon charge and $\delta$ is the curvature bias.
Simultaneous fits of $\mathcal{M}^{+}$ and $\mathcal{M}^{-}$ are performed in intervals of $\eta_{\mu}$, $\phi_{\mu}$, data-taking year, and magnet polarity to determine the pseudomass asymmetry $\mathcal{A} \equiv (\mathcal{M}^+ - \mathcal{M}^-) / (\mathcal{M}^+ + \mathcal{M}^-)$.
In turn, the curvature bias is determined from the asymmetry via $\delta = \frac{\mathcal{A}}{2}(\langle \frac{1}{p^+} \rangle + \langle \frac{1}{p^-} \rangle)$, where $\langle \frac{1}{p^\pm} \rangle$ are the mean values of the inverse of the muon momenta.

The momentum of each final-state particle is then transformed according to Eq.~\ref{eq:bias}, using the measured values of $S_\alpha$ and $\delta$.
These corrections are validated using the \ZToMuMu mass peak in data, demonstrating that they lead to improved momentum resolution. 
Given the excellent angular resolution of the \lhcb detector, negligible migrations are observed between pseudorapidity intervals in the analysis and no unfolding correction is applied.

\subsection{Corrections to the simulation}
\label{subsec:correction_sim}
To improve the description of momentum resolution in simulation, a multi-parameter smearing correction is applied to the particles' momenta as
\begin{equation}
\label{eq:smearing}
p \to p(1 + \alpha)(1 + \mathcal{R}_1\mathcal{N}\sigma_1)(1 + \mathcal{R}_2 p \sigma_2/\cosh\eta),
\end{equation}
where $\mathcal{R}_1$ and $\mathcal{R}_2$ are random variables from a standard normal distribution, $\mathcal{N}$ is a scaling factor accounting for the amount of material traversed by muon tracks up to the LHCb magnet and is set to 1.5 for muons with $\eta >$ 3.3 and unity otherwise, $\alpha$ denotes the momentum scale calibration factor, and the parameters $\sigma_1$ and $\sigma_2$ describe the momentum resolution smearing and \pt resolution effects, respectively.

\begin{figure}[!htbp]\centering
  \includegraphics[width=0.9\textwidth]{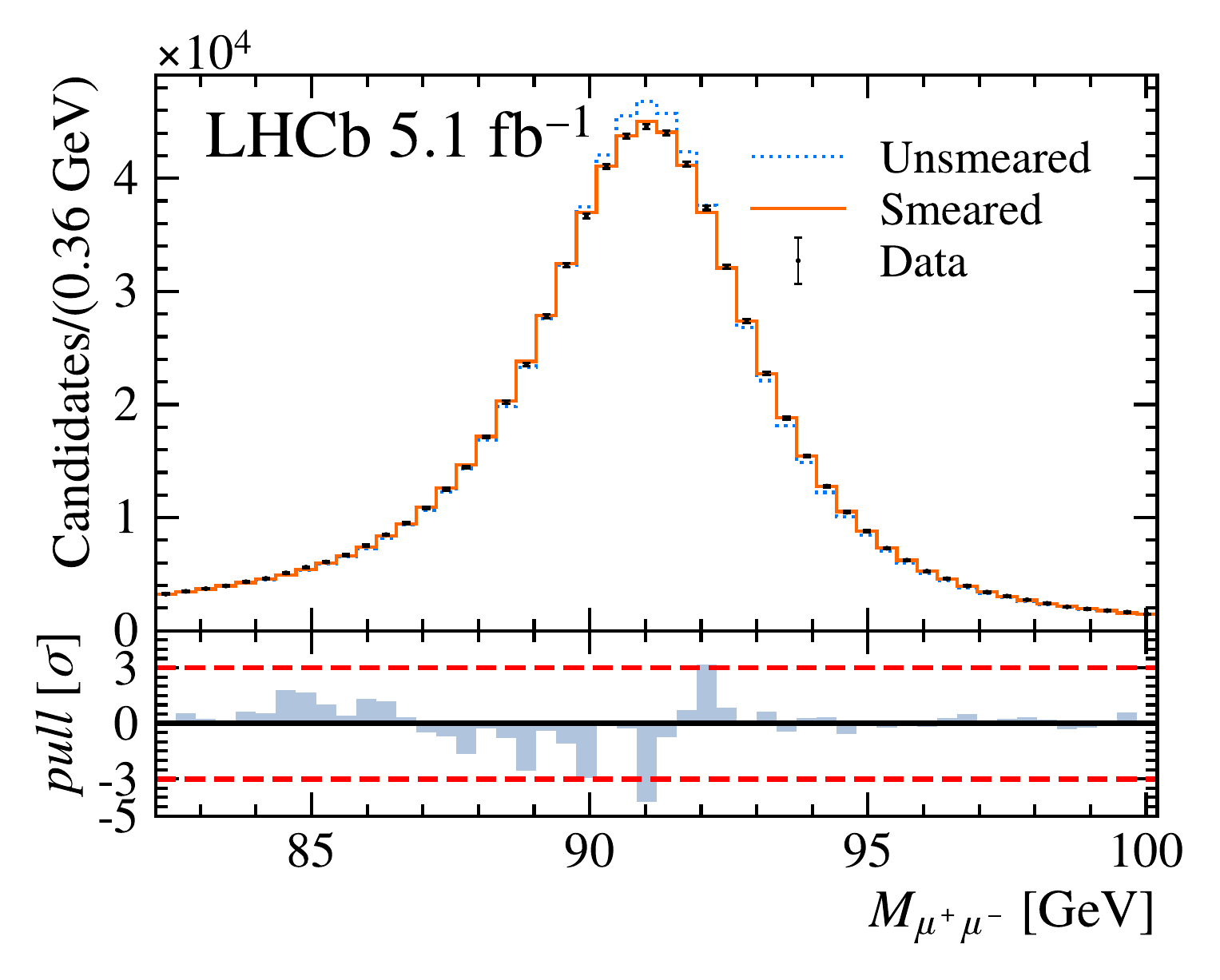}
  \caption{Invariant-mass distributions of \ZToMuMu candidates used in the momentum smearing fit, with the fit results also shown. The red and blue curves are simulated data.}
  \label{fig:MomentumScaleFit}
\end{figure}

A simultaneous maximum-likelihood fit is performed on the dimuon invariant-mass spectra of $\jpsi$, \OneS and \Z-boson candidates, incorporating both magnet-polarity configurations to extract the optimal parameters of $\alpha$, $\sigma_1$, and $\sigma_2$. The \Z-boson invariant-mass distributions with corresponding fit results are shown in Fig.~\ref{fig:MomentumScaleFit}, while numerical results including statistical uncertainties for the fitted parameters are compiled in Table~\ref{tab:MomentumScaleFit}.

\begin{table}
\centering                       \caption{\label{tab:MomentumScaleFit} Parameters of the simulation smearing model obtained from the fit procedure described in the text.}
\begin{tabular}{ccc}
\hline
&Name                                  &Value  \\
\hline
&momentum scale ($\alpha$)     &$0.128\pm0.003$ \\
&momentum smear ($\sigma_1$)     &$2.19\pm0.01$ \\
&curvature smear ($\sigma_2$)     &$15.3\pm0.5\phz$ \\
\hline
\end{tabular}
\end{table}

The default \lhcb simulation employs the \pythia generator, which lacks sufficient precision for modeling higher-order QCD effects. To account for these effects, simulation samples (\ZToMuMu, $W\rightarrow\mu\nu$ and $W\rightarrow\tau\nu$) are corrected to match NNLO predictions from the \textsc{DYTurbo} generator. Two multiplicative correction factors are calculated and applied at generator level: an angular weight ($w_{\rm ang}$) and an unpolarized cross-section weight ($w_{\rm xsec}$). The angular coefficients for each process are determined using the projection method from Ref.~\cite{Gauld:2017tww}, which computes a normalized weighted average over the dilepton angular coordinates ($\theta$, $\phi$). The method evaluates a projection function in the numerator while normalizing to the unpolarized differential cross-section in the denominator. The angular weight is then calculated as
\begin{equation}
\label{eq:AngWeight}  
w_{\rm ang} = \frac{1 + \cos^2\theta + \frac{1}{2}A_0^{\rm DYTurbo}(1 - 3\cos^2\theta) + \dots}{1 + \cos^2\theta + \frac{1}{2}A_0^{\rm Pythia}(1 - 3\cos^2\theta) + \dots},
\end{equation}
where $A$ are coefficients in the angular expansion, and the ellipses represent additional terms as defined in Ref.~\cite{Gauld:2017tww}.

For unpolarized cross-section corrections, 
simulated events are also weighted to match varying $\sigma_{\rm unpol}$ predictions using interpolation histograms. Two interpolation methods are employed: trilinear interpolation between histogram bins, and bicubic spline interpolation across parameter spaces such as $\alpha_{\rm s}$ (the strong coupling constant). This involves arrays of histograms corresponding to different parameter values. The default unpolarized cross-section weight is derived from the ratio of \textsc{DYTurbo} to \pythia predictions.

This comprehensive weighting strategy ensures proper modeling of both angular distributions and absolute cross-sections in the simulated samples.

\section{Signal yield}
\label{sec:signal_yield}

The signal yield determination employs a binned maximum-likelihood fit to the muon \pt spectra across 36 $\eta_\mu$ bins (18 intervals $\times$ 2 charges). 
The background contributions considered in this analysis include heavy-flavor processes ($\ccbar$, $\bbbar$ and $\ttbar$), electroweak processes (\ZToMuMu, \WToTauNu, Drell--Yan, $Z\to\tau^+\tau^+$, $WW$ and single top production) and QCD multijet background.

The QCD background primarily arises from hadrons produced in $pp$ collisions misidentified as muons via two mechanisms: decay-in-flight (kaons/pions decaying into muons during detector traversal) and punch-through (energetic hadrons penetrating calorimeters to reach muon detectors). These misidentification rates are parameterized as functions of momenta using simulated samples with dedicated fits for pions, kaons, and protons, as
\begin{align}
\text{pion and kaon rate} &= a_0 \, e^{\beta p} + a_1 \left(1 - e^{-\frac{m d}{\tau p}}\right) + (a_2 + a_3 \, p), \label{eqn:dif} \\
\text{proton rate} &= a_0 \, e^{\beta p} + (a_2 + a_3 \, p). \label{eqn:punch}
\end{align}
Here $a_1 \, (1 - e^{-\frac{m d}{\tau p}})$ describes the decay-in-flight probability, where $p$ is the momentum, $\tau$ the lifetime, $d$ the flight distance, and $m$ the mass of the light hadron. 
The term $(a_2 + a_3 \, p)$ models the punch-through probability as a linear function of momentum. The $a_0 \, e^{\beta p}$ functions are used to model hadrons near the edge of the LHCb detector acceptance (around $\eta = $ 2.1), where isolation requirements fail to sufficiently suppress the QCD background. Additional exponential weight factors are applied to correct for residual discrepancies in charge asymmetry and \pt distributions between data and simulation.

Furthermore, misidentification rates are parameterized as event weights in simulation. Each simulated event is tagged with its true particle type (pion, kaon, proton, or other), enabling data-driven corrections for hadron misidentification. The misidentification fit targets the three primary species (pion, kaon, proton) using the simulation, combining hadron charges to enhance statistical precision. This systematic methodology ensures robust background modeling in the subsequent analysis.

\begin{figure}[!tbp]
  \includegraphics[width=0.45\textwidth]{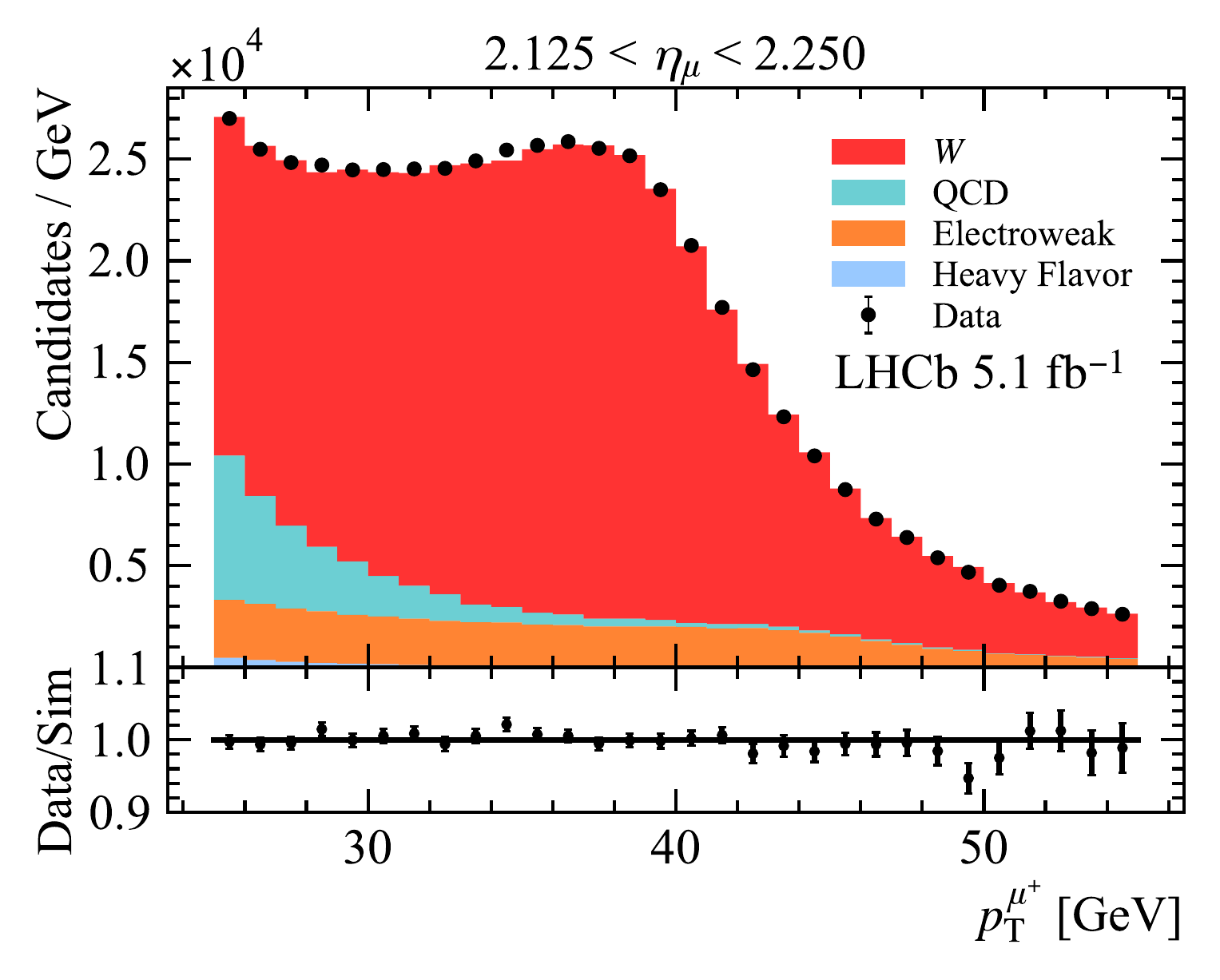}
  \includegraphics[width=0.45\textwidth]{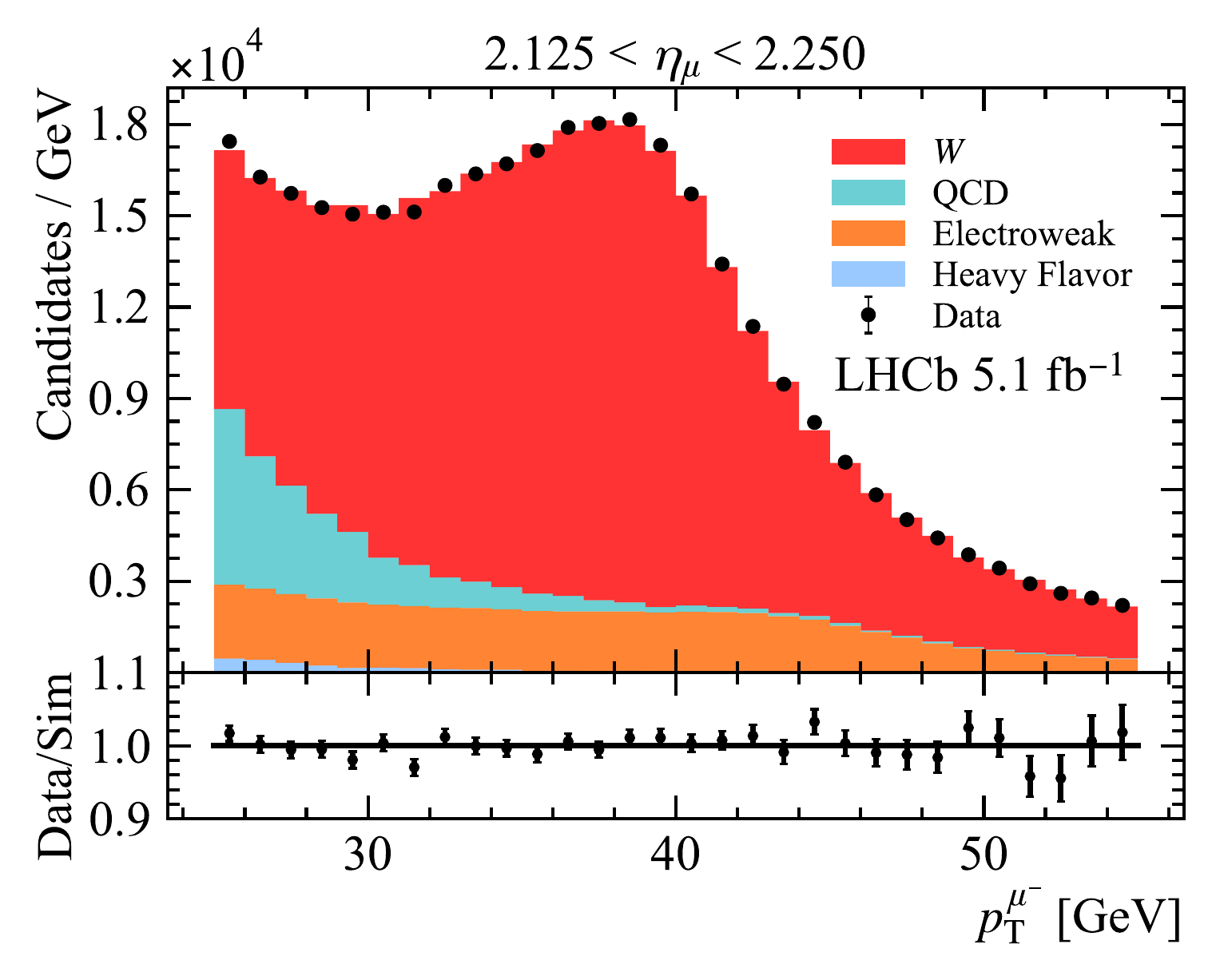}
  \caption{Distribution of the muon \pt for selected (left) \Wp and (right) \Wm bosons for the example interval 2.125 $< \eta_{\mu} <$ 2.250, with the fit results also shown.}
  \label{fig:fit_second}
\end{figure}

In the fit model, the normalization parameters for \W boson and QCD backgrounds are left free to vary, while all other background components are normalized relative to the observed number of \ZToMuMu candidates across $\eta_{\mu}$ bins via
\begin{equation}
  \label{eq:Constraint}
  f_{\rm comp} = \frac{N^{\rm obs}_Z}{N^{\rm obs}} \times \frac{\sum w}{\sum w_Z} \times \frac{\sigma^{\rm fid}}{\sigma_Z^{\rm fid}} \times \frac{N_Z^{\rm gen,fid}}{N^{\rm gen,fid}},
\end{equation}
where $f_{\rm comp}$ denotes the component fraction, $N^{\rm obs}$ the observed event yields, and $w$ corresponds to the simulation weights incorporating the efficiency. The last two factors essentially comprise a luminosity ratio relative to the \ZToMuMu simulation production, with $\sigma^{\rm fid}$ denoting fiducial cross-sections and $N^{\rm gen,fid}$ the generator-level event yields passing fiducial criteria.
The considered fiducial cross-sections are shown in Table~\ref{tab:comp_cross}. The \ZToMuMu, $Z\rightarrow\tau\tau$, $W\rightarrow\mu\nu$ and $W\rightarrow\tau\nu$ cross-sections are predicted by MCFM. Cross-sections of other components are estimated by \pythia with the \gauss~\cite{LHCb-PROC-2010-056} configuration. Example fit results for the second pseudorapidity bin are presented in Fig.~\ref{fig:fit_second}, where only statistical uncertainties are included on the data points. 

\begin{table}[!tbp]
  \centering
   \caption{Fiducial regions and cross-sections that are assumed in the signal yield fit. Here, $M$ refers to the invariant mass of the dilepton system.}
   \label{tab:comp_cross}                                                                                            \begin{tabular}{c | c | c}
   \hline
   Process                     & $\sigma^{\rm fid}$ [pb] & Fiducial region \\                                                \hline
   $Z\rightarrow\mu\mu$        & 193.5                & $p_{\rm T}^{\mu} > 20$\gev, $2.0 < \eta_{\mu} < 4.5$, $60 < M < 120$\gev\\
   $Z\rightarrow\tau\tau$      & 4.999                & $p_{\rm T}^{\tau} > 20$\gev, $2.0 < \eta_{\tau} < 4.5$, $60 < M < 120$\gev\\
   $W\rightarrow\tau\nu$       & 379.3                & $p_{\rm T}^{\tau} > 20$\gev, $2.0 < \eta_{\tau} < 4.5$\\
   $\gamma^*\rightarrow\mu\mu$ & 339.7                & $p_{\rm T}^{\mu} > 1$\gev, $2.0 < \eta_{\mu} < 4.5$, $20 < M < 40$\gev\\
   $q\overline{q}\rightarrow t\overline{t}$   & 9.259                 & $p_{\rm T}^{\mu} > 17$\gev, $2.0 < \eta_{\mu} < 4.5$\\
   $gg\rightarrow t\overline{t}$   & 36.00                 & $p_{\rm T}^{\mu} > 17$\gev, $2.0 < \eta_{\mu} < 4.5$\\
   $tW$                        & 6.723                & $p_{\rm T}^{\mu} > 17$\gev, $2.0 < \eta_{\mu} < 4.5$\\
   $WW$                        & 1.023                & $p_{\rm T}^{\mu} > 15$\gev, $2.0 < \eta_{\mu} < 4.5$\\
   $c\overline{c}$             & 2347                 & $p_{\rm T}^{\mu} > 14$\gev, $2.0 < \eta_{\mu} < 4.5$\\
   $b\overline{b}$             & $1.091\times10^{4}$            & $p_{\rm T}^{\mu} > 14$\gev, $2.0 < \eta_{\mu} < 4.5$\\
   \hline
   $W\rightarrow\mu\nu$        & 3558                 & $p_{\rm T}^{\mu} > 20$\gev, $2.0 < \eta_{\mu} < 4.5$\\
   \hline
   \end{tabular}
\end{table}

\section{Efficiencies}

The total reconstruction and selection efficiency for a given $\eta_{\mu}$ interval is defined as
\begin{equation}       
   \varepsilon_{W^{\pm}} = \frac{\sum\limits_\text{reco \W} \omega_{\rm trk,id,trg} \cdot \omega_{\rm sel} \cdot \omega_{\rm gen}}{\sum\limits_\text{true \W} \omega_{\rm gen}},
\end{equation}
where the numerator aggregates weights from events in the \W-boson simulated samples passing all selection criteria: $\omega_{\rm trk,id,trg}$ combines tracking/ID/trigger efficiencies; $\omega_{\rm sel}$ correspond to selection efficiencies; and $\omega_{\rm gen}$ denotes generator-level weights. The denominator sums generator-level weights for generated \W bosons passing fiducial requirements.

The muon reconstruction efficiency is quantified using a data-driven tag-and-probe method~\cite{LHCb-DP-2013-002} applied to \ZToMuMu decays. The efficiencies are defined as the fraction of events passing sequential selection criteria, with each stage's denominator determined by the prior stage's numerator to enable backward propagation of corrections to the fiducial selection. 
The total reconstruction efficiency is factorized as
\begin{equation} \label{eqn:eff_factorisation}
  \varepsilon = \eTrk \times \eID \times \eLZero.
\end{equation}
The tracking efficiency (\eTrk) quantifies the fraction of fiducial-acceptance events with reconstructed charged tracks meeting quality requirements; the muon identification efficiency (\eID) represents the fraction of charged tracks passing muon identification criteria; and the trigger efficiency (\eLZero) denotes the fraction of identified muon tracks satisfying the trigger selection. The efficiency is separately evaluated in each $\eta_\mu$ and $\pt^\mu$ interval with corrections applied through weighting factors that mitigate data-simulation discrepancies. Rather than bin-dependent corrections, differences are parameterized via functional fits to generate continuous weight templates.
Simulated events are further weighted using the efficiency ratio between data and simulation, $\varepsilon_{\rm data}/\varepsilon_{\rm sim}$, to correct for residual mismodeling.

\begin{figure}[!tbp]
\centering
  \includegraphics[width=0.9\textwidth]{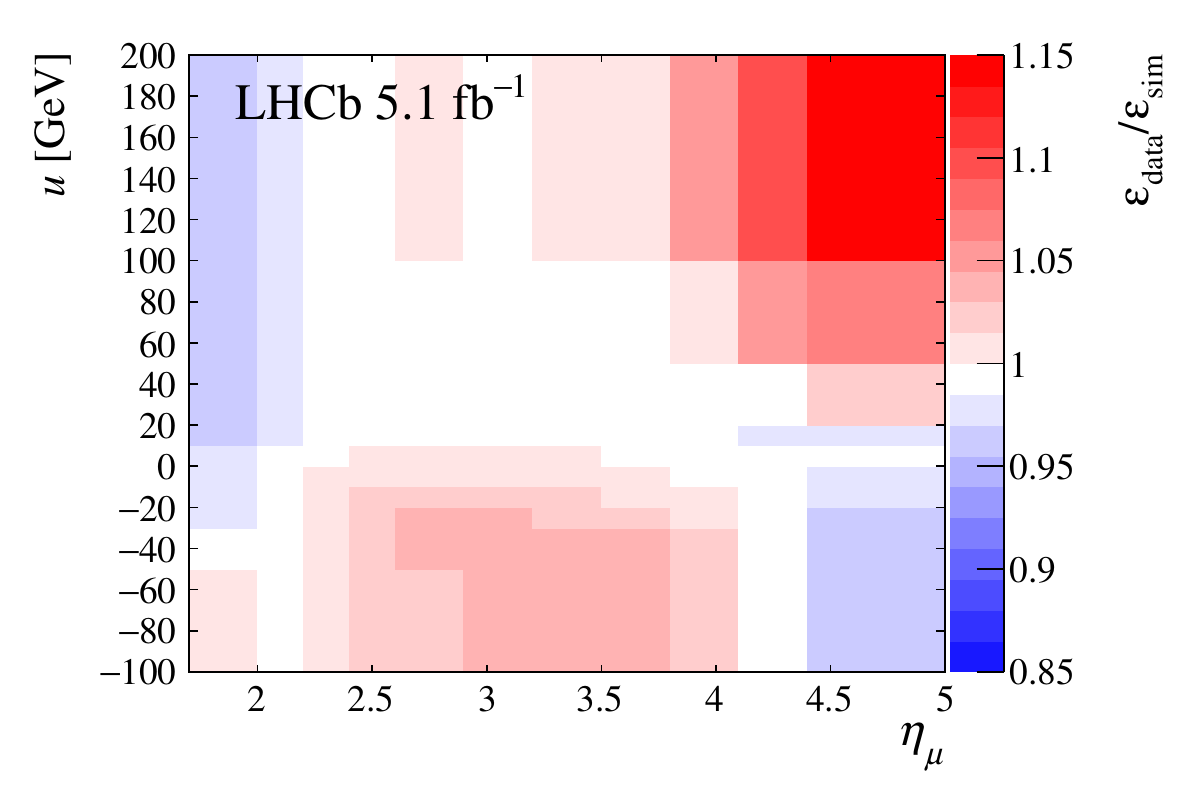}
  \caption{Data-to-simulation efficiency ratios for muon selection requirements, shown as functions of the recoil projection variable $u$ and pseudorapidity $\eta_{\mu}$. }
  \label{fig:Selection_ratio}
\end{figure}

The isolation requirement has a dominant influence on all selection criteria since its efficiency is much lower than that of other requirements. At generator level, the isolation distribution arises from two physically distinct mechanisms: underlying event activity and hard-process recoil effects. To address this complexity, a recoil projection variable is introduced for each muon, defined as
\begin{equation}
u =  \frac{\vec{p}_{\rm T}^{\,V} \cdot \vec{p}_{\rm T}^{\,\mu}} {\pt^{\mu}},
\end{equation}
where $\vec{p}^{\,\mu}_{\rm T}$ and $\vec{p}^{\,V}_{\rm T}$ respectively denote the transverse momentum vectors of the muon and its parent boson in the laboratory frame. This kinematic variable unifies selection studies across \W and \Z-boson processes, with measurable corrections extracted from \Z events as functions of reconstructed $u$. The variable effectively disentangles physical regimes: hard-process recoil effects dominate at small/negative $u$, while underlying event activity becomes dominant at large $u$. The selection efficiency exhibits greater universality when parameterized versus $u$ compared to traditional variables like $\pt^\mu$. Data-to-simulation efficiency ratios binned in $u$ and $\eta_{\mu}$ are presented in Fig.~\ref{fig:Selection_ratio}.

All efficiency corrections are incorporated as event weights in the final fit to determine signal yields, ensuring accurate consideration of reconstruction effects in the cross-section measurements. The binning schemes are designed to capture critical kinematic dependencies while preserving statistical precision in each bin. 

\section{Systematic uncertainties and stability checks}
Various sources of systematic uncertainty are investigated, with the total obtained as the quadrature sum of all components. Uncertainties in the total cross-section are compiled in Table~\ref{tab:total_unc}; those for the differential cross-sections in Tables~\ref{tab:Wp_dif_unc} and~\ref{tab:Wm_dif_unc}.

\begin{table}                   
 \centering                                  \caption{\label{tab:total_unc} Relative uncertainties affecting the \Wpm boson total cross-section measurements.} 
\begin{tabular}{cccc}
\hline
Source                         &$\Delta\sigma_{W^+\rightarrow\mu^+\nu}[\%]$ &$\Delta\sigma_{W^-\rightarrow\mu^-\overline{\nu}}[\%]$ \\
\hline
Statistical  & 0.08  & 0.11 \\
\hline
Reconstruction efficiency  & 0.33  & 0.33 \\
Generator weight  & 0.27  & 0.52 \\
QCD background  & 0.06  & 0.09 \\
Selection efficiency  & 0.51  & 0.52 \\
Smearing  & 0.02  & 0.03 \\
FSR  & 0.10  & 0.15 \\
\hline
Total Systematic (excl. lumi.)  & 0.68  & 0.83 \\
\hline
Luminosity  & 2.00  & 2.00 \\
\hline
Total  & 2.11  & 2.17 \\
\hline
\end{tabular}
\end{table}

\begin{table}
  \centering
  \caption{\label{tab:Wp_dif_unc} Summary of relative systematic uncertainties on the \Wp $\eta_{\mu}$-differential cross-section measurement from the reconstruction efficiency ($\varepsilon_{\rm rec}$), generator correction, smearing, selection efficiency ($\varepsilon_{\rm sel}$), QCD background and FSR correction, presented in percent.}
\begin{tabular}{c|cccccc}
\hline
$\eta_{\mu}$   & $\varepsilon_{\rm rec}$   & Generator    &Smearing    & $\varepsilon_{\rm sel}$   &QCDBkg   &FSR \\
\hline
2.000 -- 2.125  & 0.61 & 0.79 & 0.10 & 0.74 & 0.44 & 0.04\\
2.125 -- 2.250  & 0.58 & 0.70 & 0.05 & 0.82 & 0.22 & 0.05\\
2.250 -- 2.375  & 0.64 & 0.63 & 0.06 & 0.72 & 0.12 & 0.23\\
2.375 -- 2.500  & 0.49 & 0.58 & 0.05 & 0.70 & 0.09 & 0.03\\
2.500 -- 2.625  & 0.45 & 0.52 & 0.06 & 0.73 & 0.05 & 0.18\\
2.625 -- 2.750  & 0.42 & 0.36 & 0.04 & 0.57 & 0.05 & 0.09\\
2.750 -- 2.875  & 0.52 & 0.26 & 0.06 & 0.57 & 0.07 & 0.19\\
2.875 -- 3.000  & 0.46 & 0.25 & 0.03 & 0.49 & 0.09 & 0.17\\
3.000 -- 3.125  & 0.51 & 0.24 & 0.06 & 0.46 & 0.12 & 0.11\\
3.125 -- 3.250  & 0.41 & 0.29 & 0.05 & 0.47 & 0.12 & 0.10\\
3.250 -- 3.375  & 0.49 & 0.67 & 0.08 & 0.53 & 0.17 & 0.14\\
3.375 -- 3.500  & 0.57 & 1.17 & 0.05 & 0.53 & 0.13 & 0.13\\
3.500 -- 3.625  & 0.49 & 1.62 & 0.06 & 0.43 & 0.10 & 0.14\\
3.625 -- 3.750  & 0.53 & 1.70 & 0.07 & 0.42 & 0.13 & 0.06\\
3.750 -- 3.875  & 0.58 & 1.63 & 0.09 & 0.42 & 0.17 & 0.16\\
3.875 -- 4.000  & 0.67 & 1.48 & 0.06 & 0.44 & 0.20 & 0.28\\
4.000 -- 4.250  & 0.63 & 1.33 & 0.07 & 0.40 & 0.36 & 0.16\\
\hline
\end{tabular}
\end{table}

\begin{table}
  \centering
  \caption{\label{tab:Wm_dif_unc} Summary of relative systematic uncertainties on the \Wm $\eta_{\mu}$-differential cross-section measurement from the reconstruction efficiency ($\varepsilon_{\rm rec}$), generator correction, smearing, selection efficiency ($\varepsilon_{\rm sel}$), QCD background and FSR correction, presented in percent.}                         
\begin{tabular}{c|cccccc}
\hline
$\eta_{\mu}$   & Reco.Eff   & Generator    &Smearing    &Sel.Eff   &QCDBkg   &FSR \\
\hline
2.000 -- 2.125  & 0.59 & 0.29 & 0.07 & 0.78 & 0.51 & 0.10\\
2.125 -- 2.250  & 0.53 & 0.25 & 0.04 & 0.86 & 0.25 & 0.15\\
2.250 -- 2.375  & 0.61 & 0.24 & 0.06 & 0.75 & 0.14 & 0.18\\
2.375 -- 2.500  & 0.48 & 0.25 & 0.02 & 0.73 & 0.08 & 0.18\\
2.500 -- 2.625  & 0.45 & 0.30 & 0.03 & 0.76 & 0.06 & 0.05\\
2.625 -- 2.750  & 0.39 & 0.36 & 0.04 & 0.60 & 0.07 & 0.03\\
2.750 -- 2.875  & 0.51 & 0.35 & 0.05 & 0.60 & 0.08 & 0.16\\
2.875 -- 3.000  & 0.54 & 0.36 & 0.05 & 0.51 & 0.09 & 0.09\\
3.000 -- 3.125  & 0.54 & 0.32 & 0.05 & 0.48 & 0.11 & 0.14\\
3.125 -- 3.250  & 0.61 & 0.30 & 0.05 & 0.50 & 0.13 & 0.03\\
3.250 -- 3.375  & 0.53 & 0.30 & 0.08 & 0.55 & 0.16 & 0.26\\
3.375 -- 3.500  & 0.52 & 0.34 & 0.06 & 0.55 & 0.19 & 0.19\\
3.500 -- 3.625  & 0.59 & 0.51 & 0.07 & 0.45 & 0.17 & 0.03\\
3.625 -- 3.750  & 0.56 & 0.69 & 0.11 & 0.44 & 0.19 & 0.07\\
3.750 -- 3.875  & 0.59 & 1.09 & 0.14 & 0.43 & 0.20 & 0.27\\
3.875 -- 4.000  & 0.58 & 1.40 & 0.08 & 0.44 & 0.23 & 0.30\\
4.000 -- 4.250  & 0.67 & 2.58 & 0.10 & 0.56 & 0.30 & 0.24\\
4.250 -- 4.500  & 0.94 & 4.75 & 0.31 & 0.75 & 0.52 & 0.39\\
\hline
\end{tabular}
\end{table}

Three sources contribute to the systematic uncertainty from the reconstruction efficiency: the statistical uncertainty from finite tag-and-probe sample sizes, assessed via Gaussian fluctuations of $\varepsilon_{\rm sim}$ and $\varepsilon_{\rm data}/\varepsilon_{\rm sim}$ within uncertainties before efficiency fitting; alternative parameterizations for tracking/ID/trigger efficiencies; and variations in denominator selection criteria.

Generator-level systematic uncertainties comprise three components: scale variations quantified via the envelope of 7-point scale variations for angular coefficients and unpolarized cross-sections; the unpolarized configuration uncertainty determined by comparing \textsc{DYTurbo} \vs \powheg+ \pythia (NLO) predictions with/without unpolarized weights; and PDF uncertainties probed using NNPDF31\_nnlo\_as\_0118~\cite{NNPDF:2017mvq}, MSHT20nnlo\_as118~\cite{Bailey:2020ooq}, and CT18NNLO~\cite{Hou:2019efy} sets, with the envelope defining the systematic uncertainty. The associated uncertainty for \Wm is much larger than for \Wp, which is driven by the contribution from the highest $\eta_\mu$ interval (see Tables~\ref{tab:Wp_dif_unc}--\ref{tab:Wm_dif_unc}).

Systematic uncertainties in QCD background estimation comprise two components: the statistical uncertainty in data/simulation and misidentification fit parameters, assessed via multivariate Gaussian resampling; and the assumed pion, kaon, and proton abundances, quantified by repeating the measurement with these varied independently by $\pm 20\%$.

The selection efficiency systematic uncertainty has three components:
the RMS (Root Mean Square) of cross-section variations derived from repeating efficiency corrections with Gaussian-smeared data/simulation efficiency ratios per bin; the envelope of the cross-section variations from alternative binning schemes for Fig.~\ref{fig:Selection_ratio}; and the cross-section variation due to the tightening of the \Z-boson mass window of the denominator from $60< M <120\gev$ to $70< M < 110\gev$.

The systematic uncertainty from track smearing comprises two components: the statistical uncertainty in smearing model parameters, assessed via multivariate Gaussian resampling of parameter correlations from the covariance matrix of the fit; and the statistical uncertainty on the curvature bias $\delta$.

To estimate the systematic uncertainty from the treatment of FSR, the \powheg generator is used, where the FSR is modeled using \pythia. The differences of FSR corrections between \resbos + \photos and \powheg + \pythia are assigned as the systematic uncertainty.

In addition to the experimental systematic uncertainties, an overall cross-section uncertainty of 2.0\% is assigned to account for the limited precision on the integrated luminosity of the sample~\cite{LHCb-PAPER-2014-047}. This uncertainty is 100\% correlated between bins in the differential measurements.

To validate the result robustness, the full analysis has been repeated in subsamples obtained partitioning by year and magnet polarity, with all results being consistent. Alternative fits include those with: an alternative pt binning scheme; the \WToTauNu cross-section varied by $\pm 1$\%; the \ccbar/\bbbar cross-sections varied by $+50$\%; the efficiency correction split into two independent $\phi$ regions; and an alternative QCD background simulation with the \pt threshold reduced from 18 to 14\gev. The resulting variations are consistent with statistical fluctuations.

\section{Results}
\begin{figure}[!tbp]
\centering      
\includegraphics[width=0.9\textwidth]{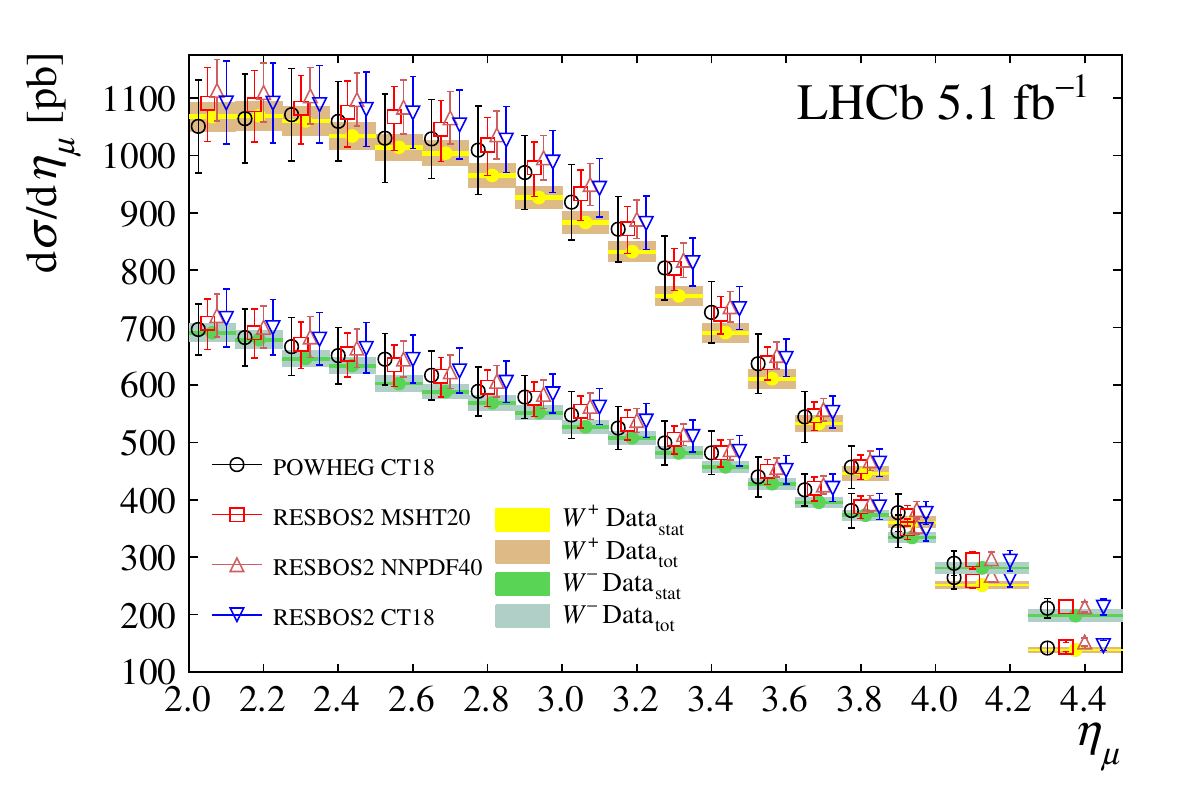}
  \caption{Differential \WToMuNu cross-sections, as a function of $\eta_\mu$. The bands represent the data, with the inner region corresponding to the statistical uncertainty and the outer one to the total uncertainty. For theoretical predictions, the error bars incorporate both PDF and scale uncertainties.}
  \label{fig:dif_cross_result}
\end{figure}

\begin{figure}[!tbp]
\centering
 \includegraphics[width=0.7\textwidth]{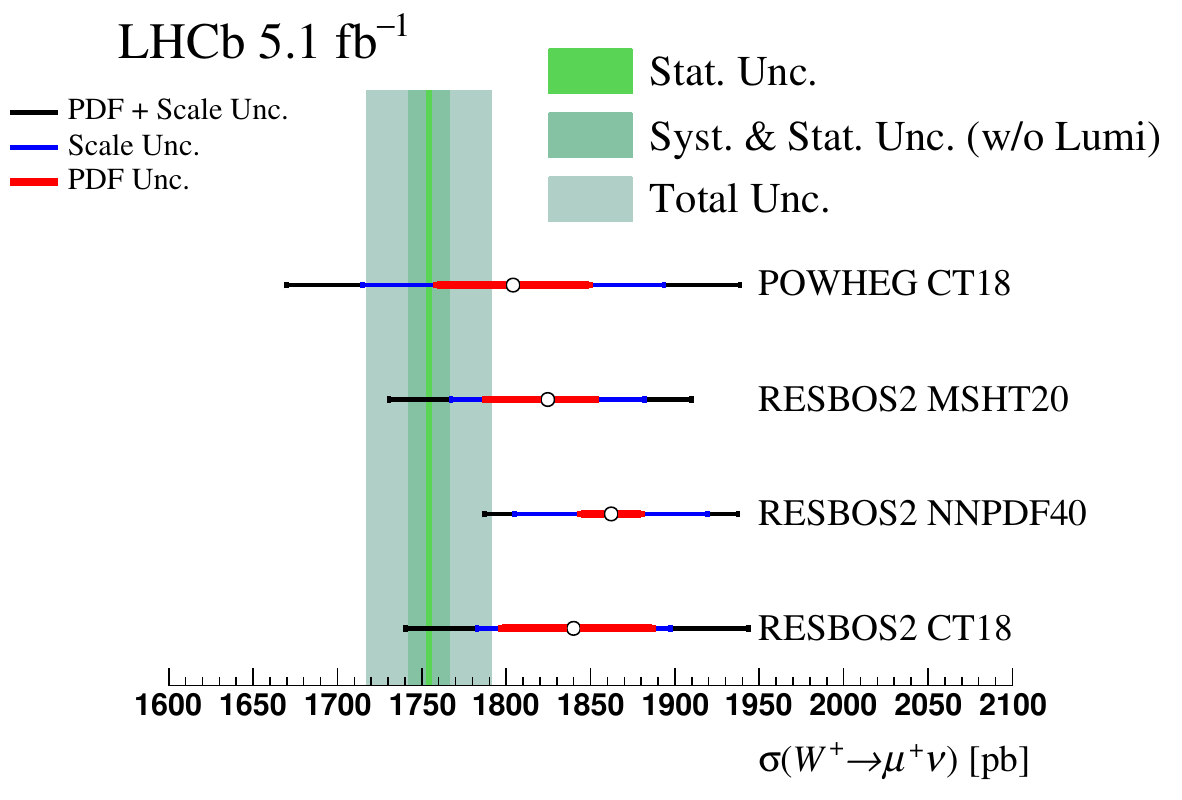}
 \includegraphics[width=0.7\textwidth]{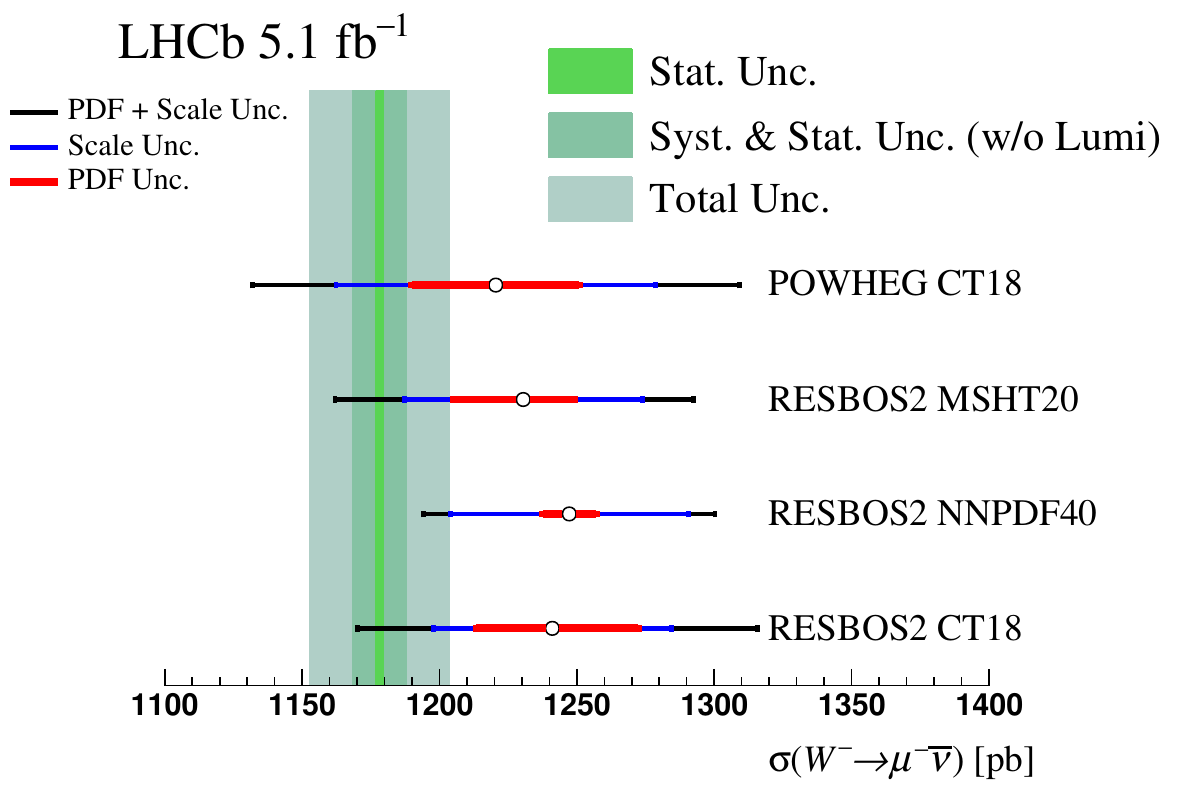} 
\caption{Total (top) \WpToMuNu and (bottom) \WmToMuNu cross-sections compared with theoretical predictions. The bands denote the measurements, with the inner region corresponding to the statistical and the outer one to the total uncertainty. For the theoretical predictions, the error bars incorporate both PDF and scale uncertainties.}
  \label{fig:cross_result}
\end{figure}

The measured cross-section results are computed from Eq.~\ref{eq:cross-section} and presented in Table~\ref{tab:result}, Fig.~\ref{fig:dif_cross_result}, and Fig.~\ref{fig:cross_result}. Theoretical predictions employ NNLO \resbos2 coupled to CT18, NNPDF40~\cite{NNPDF:2021njg}, and MSHT20 PDF sets; additional comparisons use NLO \powheg + \pythia with CT18 configurations.

\begin{table}[!tbp]
  \centering
  \caption{\label{tab:result} Measured differential \WToMuNu cross-sections, as a function of muon pseudorapidity. The first uncertainty is statistical, the second systematic, and the third due to the luminosity (which is 100\% correlated between bins). }
\begin{tabular}{ccc}
\hline
$\eta_{\mu}$   & $\deriv\sigma_{W^+} / \deriv\eta_{\mu}$ [pb]   & $\deriv\sigma_{W^-} / \deriv\eta_{\mu}$ [pb] \\
\hline
2.000 -- 2.125    & 1067.6 $\pm$ 3.2  $\pm$ 14.1  $\pm$ 21.4        &691.0 $\pm$ 2.7 $\pm$ 7.9 $\pm$ 13.8\\
2.125 -- 2.250   & 1069.3 $\pm$ 3.0 $\pm$ 13.3 $\pm$ 21.4       &679.1 $\pm$ 2.6 $\pm$ 7.3 $\pm$ 13.6\\
2.250 -- 2.375   & 1060.3 $\pm$ 3.3 $\pm$ 12.5 $\pm$ 21.2       &646.2 $\pm$ 2.5 $\pm$ 6.6 $\pm$ 12.9\\
2.375 -- 2.500    & 1034.0 $\pm$ 3.2  $\pm$ 10.7  $\pm$ 20.7        &633.8 $\pm$ 2.5 $\pm$ 5.9 $\pm$ 12.7\\
2.500 -- 2.625    & 1014.3 $\pm$ 3.2  $\pm$ 10.4  $\pm$ 20.3        &602.5 $\pm$ 2.4 $\pm$ 5.6 $\pm$ 12.1\\
2.625 -- 2.750   & 1004.0 $\pm$ 3.1 $\pm$ \phz8.1  $\pm$ 20.1        &588.4 $\pm$ 2.4 $\pm$ 4.7 $\pm$ 11.8\\
2.750 -- 2.875   & \phz965.2 $\pm$ 3.0  $\pm$ \phz8.1   $\pm$ 19.3         &568.8 $\pm$ 2.3 $\pm$ 5.0 $\pm$ 11.4\\
2.875 -- 3.000    & \phz926.7 $\pm$ 3.1   $\pm$ \phz6.9    $\pm$ 18.5          &551.7 $\pm$ 2.4 $\pm$ 4.6 $\pm$ 11.0\\
3.000 -- 3.125    & \phz883.2 $\pm$ 3.1   $\pm$ \phz6.6    $\pm$ 17.7          &527.0 $\pm$ 2.4 $\pm$ 4.3 $\pm$ 10.5\\
3.125 -- 3.250   & \phz832.4 $\pm$ 2.9  $\pm$ \phz5.9   $\pm$ 16.6         &508.0 $\pm$ 2.1 $\pm$ 4.3 $\pm$ 10.2\\
3.250 -- 3.375   & \phz755.3 $\pm$ 2.7  $\pm$ \phz7.6   $\pm$ 15.1         &481.7 $\pm$ 2.1 $\pm$ 4.2 $\pm$ \phz9.6\\
3.375 -- 3.500    & \phz691.2 $\pm$ 2.4   $\pm$ \phz9.8    $\pm$ 13.8          &457.4 $\pm$ 2.1 $\pm$ 4.0 $\pm$ \phz9.1\\
3.500 -- 3.625    & \phz610.8 $\pm$ 2.4   $\pm$ 10.7   $\pm$ 12.2         &428.0 $\pm$ 2.2 $\pm$ 3.9 $\pm$ \phz8.6\\
3.625 -- 3.750   & \phz532.8 $\pm$ 2.3  $\pm$ \phz9.8   $\pm$ 10.7         &395.7 $\pm$ 2.1 $\pm$ 4.0 $\pm$ \phz7.9\\
3.750 -- 3.875   & \phz446.0 $\pm$ 2.0  $\pm$ \phz8.0   $\pm$ \phz8.9          &373.0 $\pm$ 2.2 $\pm$ 5.1 $\pm$ \phz7.5\\
3.875 -- 4.000    & \phz361.1 $\pm$ 1.7   $\pm$ \phz6.2    $\pm$ \phz7.2           &334.4 $\pm$ 2.0 $\pm$ 5.4 $\pm$ \phz6.7\\
4.000 -- 4.250     & \phz251.3 $\pm$ 1.2    $\pm$ \phz4.0     $\pm$ \phz5.0            &281.2 $\pm$ 1.3 $\pm$ 7.7 $\pm$ \phz5.6\\
4.250 -- 4.500     & \phz138.3 $\pm$ 1.2    $\pm$ \phz3.7     $\pm$ \phz2.8            &197.9 $\pm$ 1.6 $\pm$ 9.8 $\pm$ \phz4.0\\
\hline
\end{tabular}
\end{table}

The integrated production cross-sections of $W$-boson and the ratios of \W to \Z-boson total cross-sections~\cite{LHCb-PAPER-2021-037} are determined to be
\begin{equation*}
\begin{aligned}
  \sigma_{W^+ \to \mu^+\nu} &= 1754.2 \pm 1.5 \pm 11.9 \pm 35.1\text{ pb}\\
  \sigma_{W^- \to \mu^-\overline{\nu}} &= 1178.1 \pm 1.3 \pm \phz9.7 \pm 23.6\text{ pb}\\
\end{aligned}
\end{equation*} 

\begin{equation*}
\begin{aligned}
  R_{W^+Z} \equiv \frac{\sigma_{W^+ \to \mu^+\nu}}{\sigma_{Z \to \mu^+\mu^-}} = 8.932 \pm 0.012 \pm 0.081\text{,} \\
  R_{W^-Z} \equiv \frac{\sigma_{W^- \to \mu^-\overline{\nu}}}{\sigma_{Z \to \mu^+\mu^-}} = 5.999 \pm 0.009 \pm 0.061\text{,} \\
  R_{WZ} \equiv \frac{\sigma_{W \to \mu\nu}}{\sigma_{Z \to \mu^+\mu^-}} = 14.930 \pm 0.026 \pm 0.134\text{,} \\
\end{aligned}
\end{equation*} 
where the uncertainties are statistical, systematic, and where relevant, due to the luminosity determination, respectively.

\begin{figure}[!tbp]
  \includegraphics[width=0.90\textwidth]{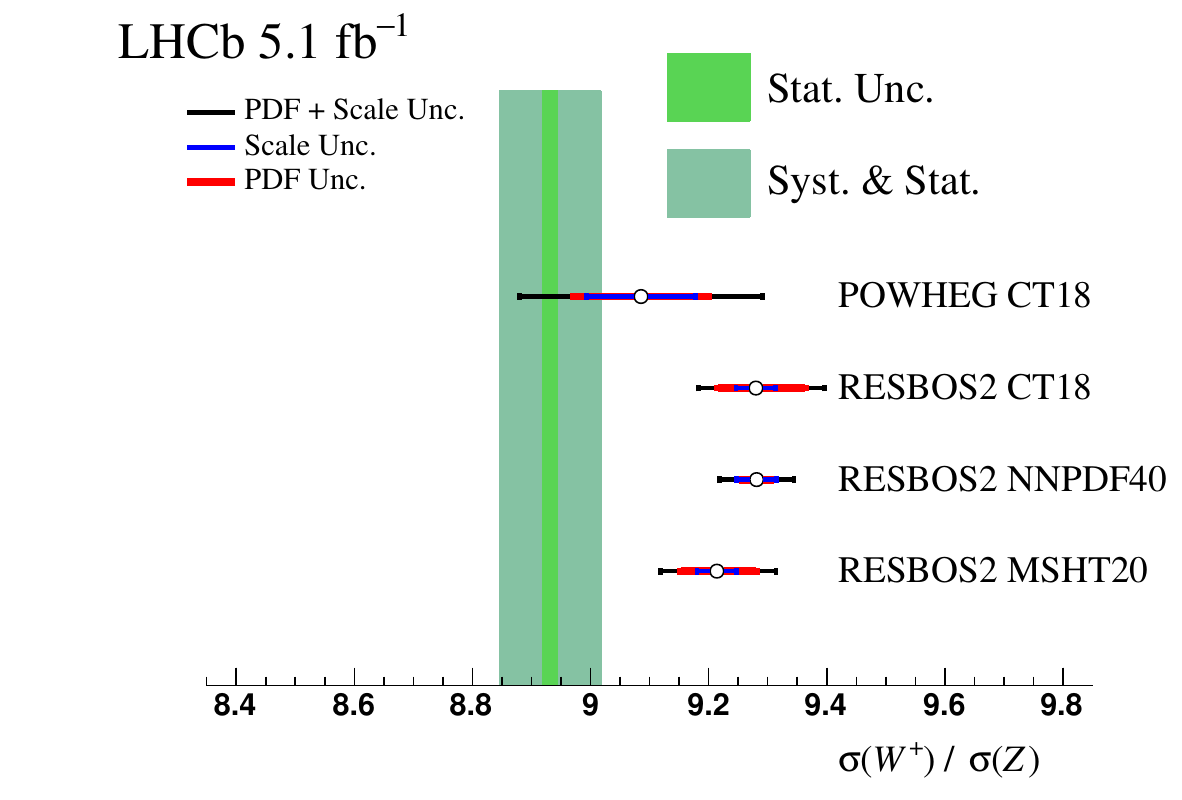}\\
  \includegraphics[width=0.90\textwidth]{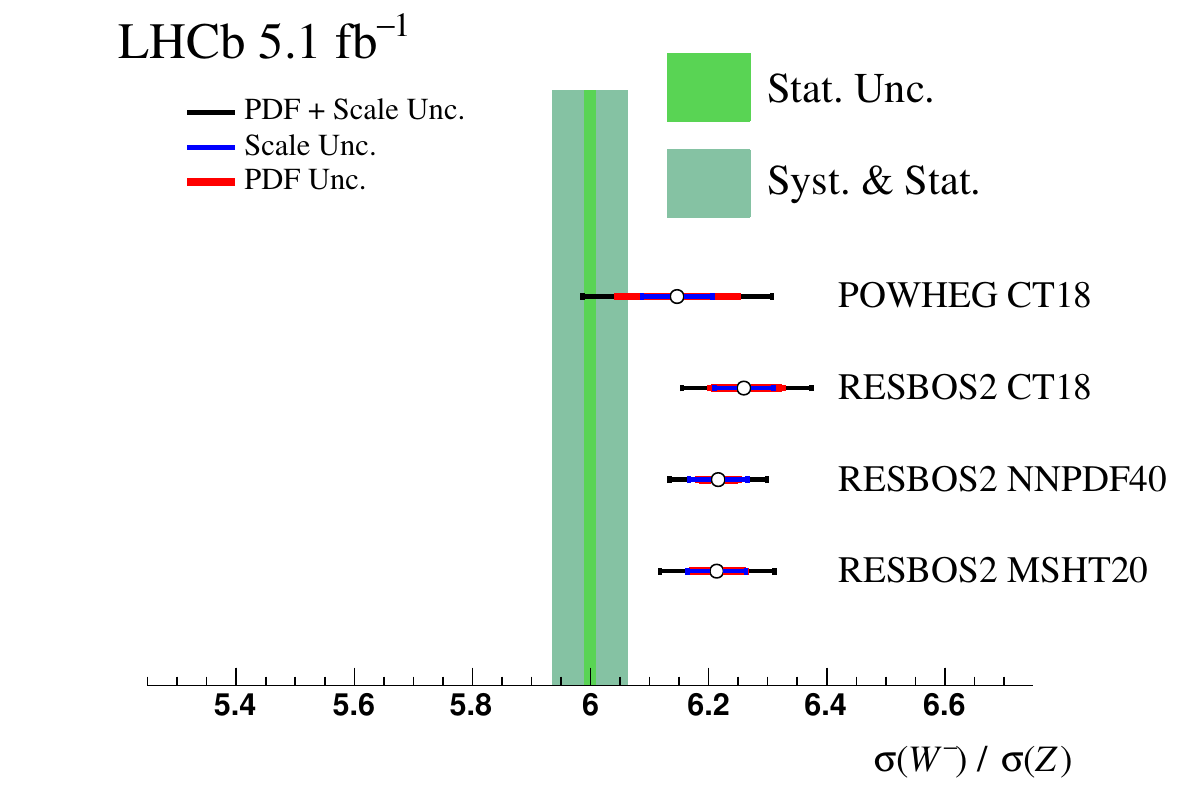}
  \caption{Results for the (top) $\sigma_{W^+ \to \mu^+\nu} / \sigma_{Z \to \mu^+\mu^-}$ and (bottom) $\sigma_{W^- \to \mu^-\overline{\nu}} / \sigma_{Z \to \mu^+\mu^-}$ cross-section ratios compared with theoretical predictions. The $\pt^{\mu}$ range used for the $W$ ($Z$) cross-section measurement is 25 $< \pt^{\mu} <$ 55\gev ($\pt^{\mu} > $ 20\gev). The bands denote the measurements, with the inner region corresponding to the statistical uncertainty and the outer region corresponding to the statistical and systematic uncertainty. The theoretical points correspond to the different predictions, with error bars including both PDF and scale uncertainties.}
  \label{fig:W2Z_ratio}
\end{figure}

\begin{figure}[!tbp]\centering
  \includegraphics[width=0.55\textwidth]{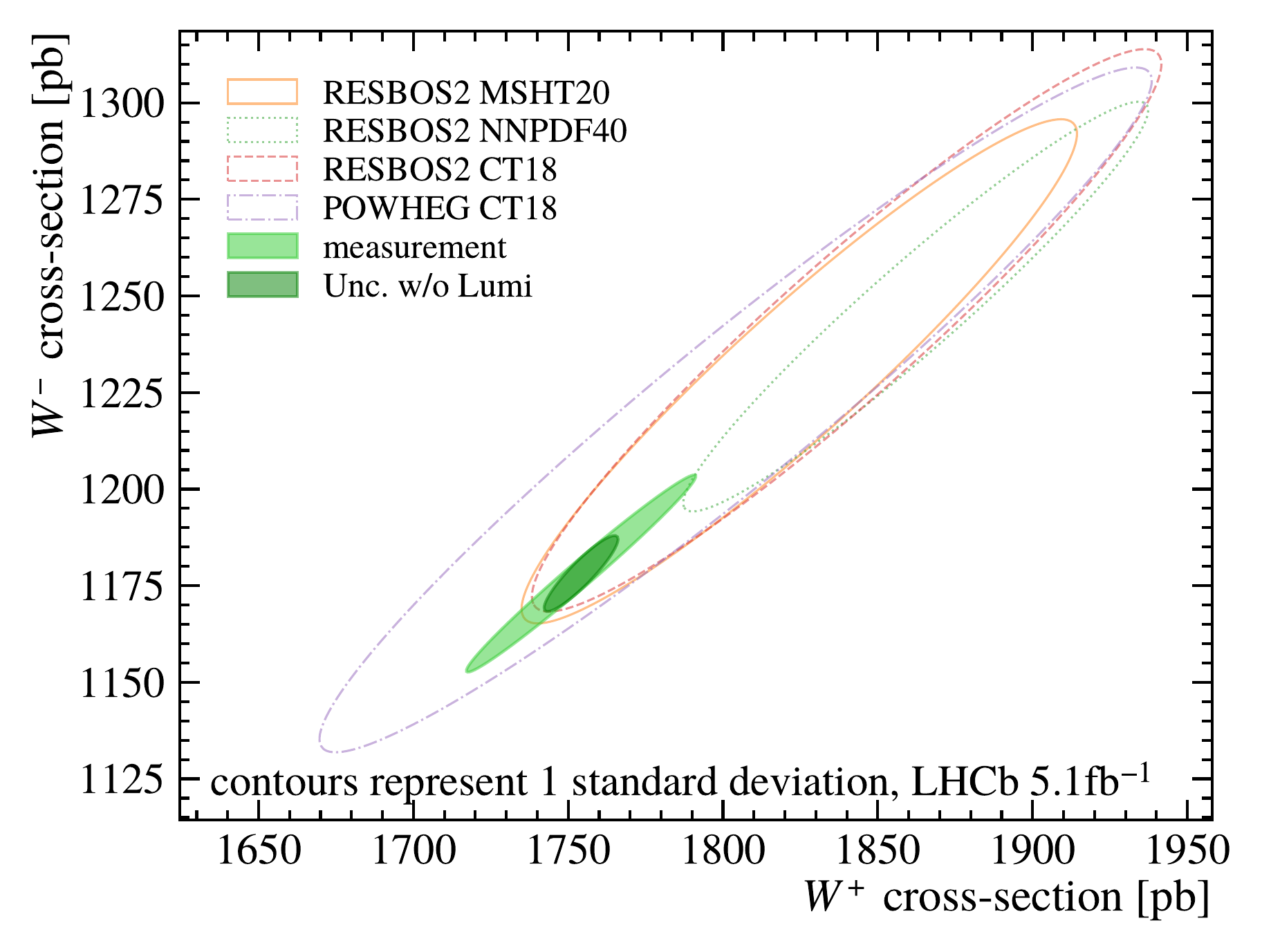}\\
  \includegraphics[width=0.55\textwidth]{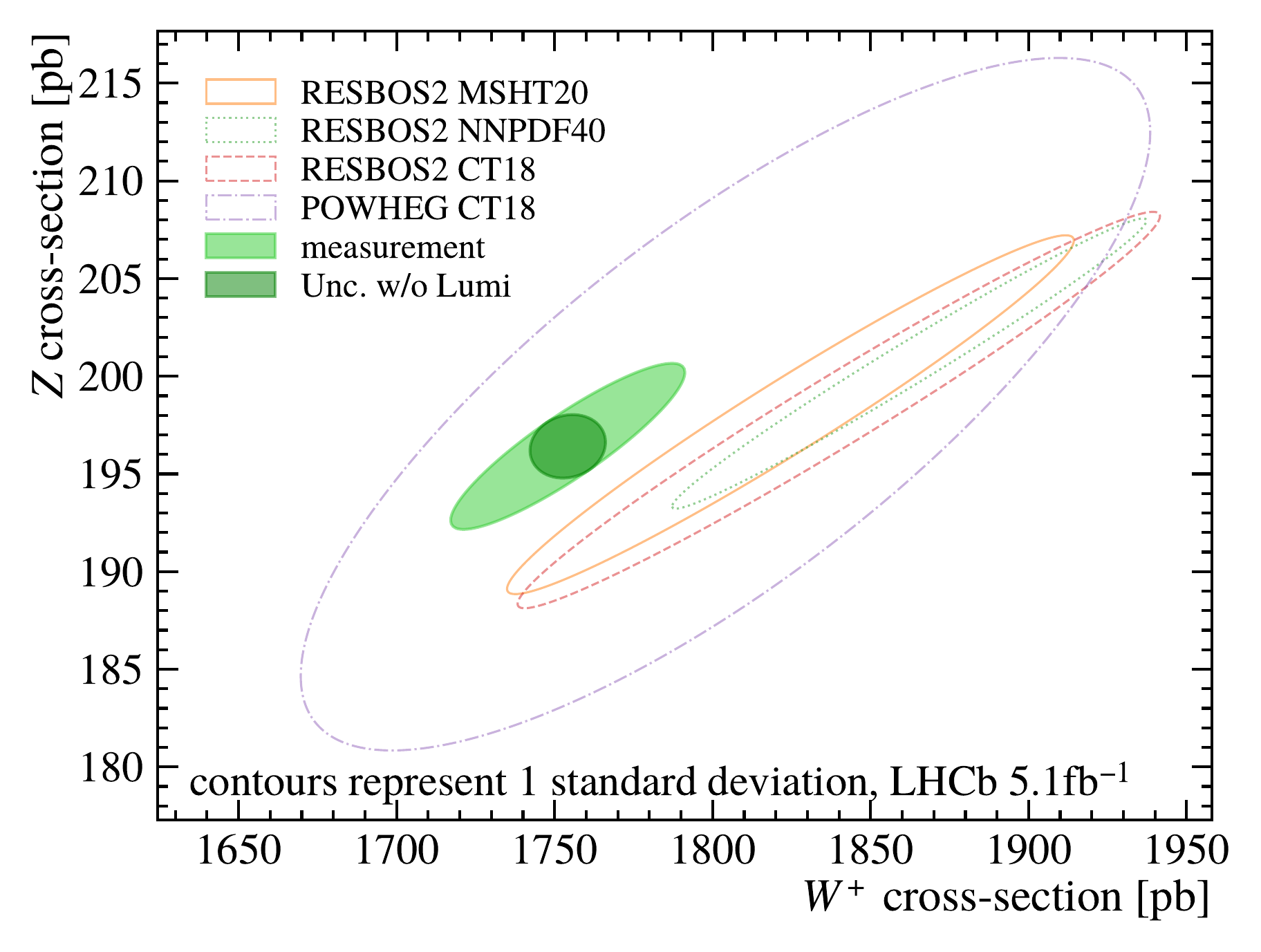}\\
  \includegraphics[width=0.55\textwidth]{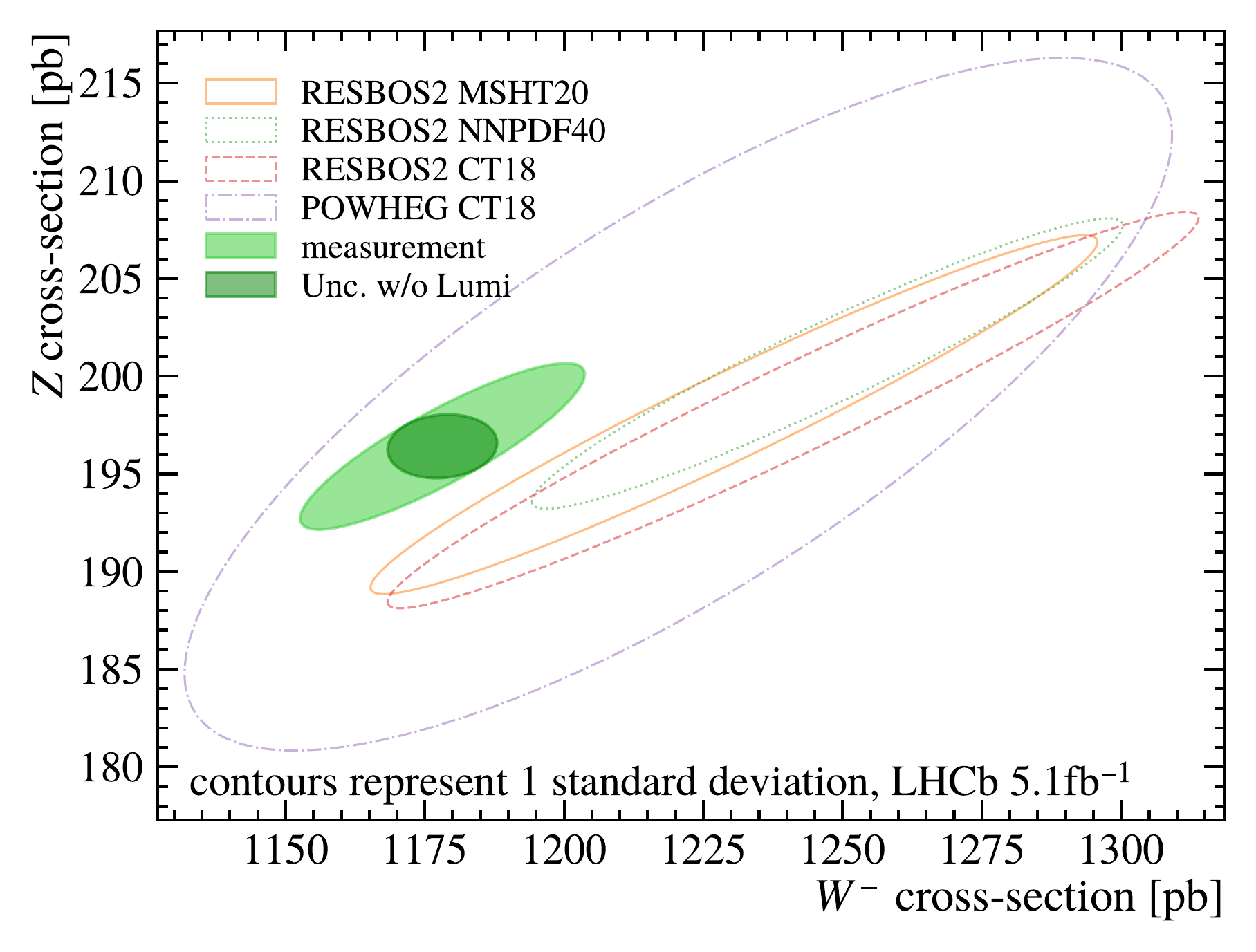}\\
  \caption{Integrated cross-sections measurements of (top) $W^+$ and $W^-$, (middle) $W^+$ and $Z$, and (bottom) $W^-$ and $Z$, compared with theoretical predictions. The $\pt^{\mu}$ range used for the $W$ ($Z$) cross-sections measurement is 25 $< \pt^{\mu} <$ 55\gev ($\pt^{\mu} > $ 20\gev). The inner ellipse of the measurement includes statistical and systematic uncertainties, but excludes the luminosity uncertainty. }
  \label{fig:ellipse}
\end{figure}

In determining these ratios, all uncertainties are assumed to be uncorrelated, with the exception of the systematic uncertainty relating to the reconstruction efficiency, where a 50\% positive correlation is assumed. The results are compared to theoretical predictions in Fig.~\ref{fig:W2Z_ratio}, while the $Z$ and $W$-boson integrated cross-section correlations are shown in Fig.~\ref{fig:ellipse}.

\section{Conclusion}
Precision measurements of \Wpm-boson production in proton-proton collisions at \mbox{$\sqs=13\tev$} are presented using muon decay channels. The analysis utilizes 5.1\invfb of data collected by the \lhcb detector during Run 2 (2016--2018). Inclusive \WToMuNu production cross-sections are measured within the fiducial region defined by muon transverse momentum $25 < \pt^{\mu} < 55\gev$ and pseudorapidity $2.0 < \eta_{\mu} < 4.5$.

The measured differential cross-sections exhibit excellent consistency with NNLO perturbative QCD predictions calculated via \resbos2 using CT18 PDFs. These high-precision measurements impose stringent constraints on parton distribution functions, particularly in sparsely constrained small-$x$ and large-$x$ regions, with experimental uncertainties matching current theoretical precision. The findings underscore the unique capability of forward muon measurements at the \lhcb experiment to probe PDFs in kinematic regimes inaccessible to other general-purpose detectors.

%% file: LHCb/acknowledgements.tex
\section*{Acknowledgements}
%
%
\noindent We express our gratitude to our colleagues in the CERN
accelerator departments for the excellent performance of the LHC. We
thank the technical and administrative staff at the LHCb
institutes.
We acknowledge support from CERN and from the national agencies:
ARC (Australia);
CAPES, CNPq, FAPERJ and FINEP (Brazil); 
MOST and NSFC (China); 
CNRS/IN2P3 and CEA (France);  
BMFTR, DFG and MPG (Germany);
INFN (Italy); 
NWO (Netherlands); 
MNiSW and NCN (Poland); 
MEC/IFA (Romania); 
MICIU and AEI (Spain);
SNSF and SER (Switzerland); 
NASU (Ukraine); 
STFC (United Kingdom); 
DOE NP and NSF (USA).
We acknowledge the computing resources that are provided by ARDC (Australia), 
CBPF (Brazil),
CERN, 
IHEP and LZU (China),
IN2P3 (France), 
KIT and DESY (Germany), 
INFN (Italy), 
SURF (Netherlands),
Polish WLCG (Poland),
IFIN-HH (Romania), 
PIC (Spain), CSCS (Switzerland), 
GridPP (United Kingdom),
and NSF (USA).  
We are indebted to the communities behind the multiple open-source
software packages on which we depend.
Individual groups or members have received support from
RTP (Australia), 
Key Research Program of Frontier Sciences of CAS, CAS PIFI, CAS CCEPP (China); 
Minciencias (Colombia);
EPLANET, Marie Sk\l{}odowska-Curie Actions, ERC and NextGenerationEU (European Union);
A*MIDEX, ANR, IPhU and Labex P2IO, and R\'{e}gion Auvergne-Rh\^{o}ne-Alpes (France);
Alexander-von-Humboldt Foundation (Germany);
ICSC (Italy); 
Severo Ochoa and Mar\'ia de Maeztu Units of Excellence, GVA, XuntaGal, GENCAT, InTalent-Inditex and Prog.~Atracci\'on Talento CM (Spain);
the Leverhulme Trust, the Royal Society and UKRI (United Kingdom).

%% file: Authorship_LHCb-PAPER-2025-070.tex
\centerline
{\large\bf LHCb collaboration}
\begin
{flushleft}
\small
R.~Aaij$^{38}$\lhcborcid{0000-0003-0533-1952},
M. ~Abdelfatah$^{69}$,
A.S.W.~Abdelmotteleb$^{57}$\lhcborcid{0000-0001-7905-0542},
C.~Abellan~Beteta$^{51}$\lhcborcid{0009-0009-0869-6798},
F.~Abudin{\'e}n$^{59}$\lhcborcid{0000-0002-6737-3528},
T.~Ackernley$^{61}$\lhcborcid{0000-0002-5951-3498},
A. A. ~Adefisoye$^{69}$\lhcborcid{0000-0003-2448-1550},
B.~Adeva$^{47}$\lhcborcid{0000-0001-9756-3712},
M.~Adinolfi$^{55}$\lhcborcid{0000-0002-1326-1264},
P.~Adlarson$^{87}$\lhcborcid{0000-0001-6280-3851},
C.~Agapopoulou$^{14}$\lhcborcid{0000-0002-2368-0147}
% ...

{\footnotesize \it

% ...
$^{14}$Universit{\'e} Paris-Saclay, CNRS/IN2P3, IJCLab, Orsay, France\\
% ...
$^{38}$Nikhef National Institute for Subatomic Physics, Amsterdam, Netherlands\\
% ...
$^{47}$Instituto Galego de F{\'\i}sica de Altas Enerx{\'\i}as (IGFAE), Universidade de Santiago de Compostela, Santiago de Compostela, Spain\\
% ...
$^{51}$Physik-Institut, Universit{\"a}t Z{\"u}rich, Z{\"u}rich, Switzerland\\
% ...
$^{55}$H.H. Wills Physics Laboratory, University of Bristol, Bristol, United Kingdom\\
% ...
$^{57}$Department of Physics, University of Warwick, Coventry, United Kingdom\\
% ...
$^{59}$School of Physics and Astronomy, University of Edinburgh, Edinburgh, United Kingdom\\
$^{60}$School of Physics and Astronomy, University of Glasgow, Glasgow, United Kingdom\\
$^{61}$Oliver Lodge Laboratory, University of Liverpool, Liverpool, United Kingdom\\
% ...
$^{69}$Syracuse University, Syracuse, NY, United States\\
% ...
$^{87}$Department of Physics and Astronomy, Uppsala University, Uppsala, Sweden, associated to $^{60}$\\
% ...
\medskip
}
\end{flushleft}